\documentclass[prb,aps,twocolumn,eqsecnum,superscriptaddress]{revtex4}
\usepackage{graphicx}
\usepackage{subfigure}
\usepackage{pstricks}
\usepackage{amssymb}
\usepackage{amsmath}
\newlength{\myL}

\newcommand{\rw}{\rightarrow}
\newcommand{\pd}{\partial}
\newcommand{\cC}{\mathcal{C}}

\newcommand{\si}{\sigma}

\newcommand{\Tr}{{\rm Tr}}

\def\ket#1{\left|{#1}\right>}
\def\bra#1{\left<{#1}\right|}
\def\braa#1{\left<{#1}\right.}

\begin{document}

\title{Topological phases and topological entropy of two-dimensional systems with finite correlation length}

\author{Stefanos Papanikolaou}
\affiliation{Department of Physics, University of Illinois at
Urbana-Champaign, 1110 W. Green St., Urbana, IL  61801-3080}
\author{Kumar S. Raman}
\affiliation{Department of Physics and Astronomy, University of California at Riverside, Riverside, CA 92521}
\author{Eduardo Fradkin}
\affiliation{Department of Physics, University of Illinois at
Urbana-Champaign, 1110 W. Green St., Urbana, IL  61801-3080}

\date{\today}

\begin{abstract}
We elucidate the topological features of the entanglement entropy of a region in two dimensional quantum systems in a topological phase with a finite correlation length $\xi$. Firstly, we suggest that simpler reduced quantities, related to the von Neumann entropy, could be defined to compute the topological entropy. We use our methods to compute the entanglement entropy for the ground state wave function of a quantum eight-vertex model in its topological phase, and show that a finite correlation length adds corrections of the same order as  the topological entropy which come from sharp features of the boundary of the region under study. We also calculate the topological entropy for the ground state of the  quantum dimer model on a triangular lattice by using a mapping to a loop model. The topological entropy of the state is determined by loop configurations with a non-trivial winding number around the region under study. Finally, we consider 
extensions of the Kitaev wave function, which incorporate the effects of electric and magnetic charge fluctuations, and use it to investigate the stability of the topological phase by calculating the topological entropy.
\end{abstract}


\maketitle
\section{Introduction}

Different phases of condensed matter are usually distinguished by broken symmetries of the physical system measured by local order parameters, an approach that is ultimately justified by the renormalization group.  In quantum systems at zero temperature, these local order parameters correspond to local observables having nonzero expectation value in the ground state wave function in the broken symmetry phase.  However, quantum systems can also display phases where the character of the state is not measurable by any local order parameter.  An example is topological order\cite{wen90a, wen95}, a form of quantum order marked by a sensitivity of the ground state degeneracy to the topology of the system in the absence of any broken symmetry.  An experimental system with topological order is the two-dimensional electron gas  in the regime of the fractional quantum Hall effect.\cite{wen90a} A number of simple and very idealized theoretical models\cite{kitaev97,moessner2001,levin05,Fendley05} (not yet realized in experiment) with topological phases have now been constructed.

Recently, Kitaev and Preskill\cite{kitaev2006} and Levin and Wen\cite{levin2006} proposed that, in two-dimensional systems, whether a ground state wave function $|\Psi\rangle$ has topological order can be determined by computing a nonlocal quantity called the topological entanglement entropy $-\gamma$.  The computation involves dividing the system into two subregions, $A$ and $B$; obtaining the reduced density matrix $\rho_A = Tr_B |\Psi\rangle\langle\Psi|$ by tracing over the degrees of freedom in $B$; and then calculating the von Neumann entropy $S_A = -Tr_A \rho_A \ln \rho_A$ by tracing over region $A$.  The observation was that if the boundary between regions $A$ and $B$ has perimeter $L$ and is sufficiently smooth, then for a system with short range correlations, the entropy should have the scaling form:
\begin{equation}
S_A = \alpha L - \gamma + \dots
\label{eq:sA}
\end{equation}
where $\alpha$ is a non-universal finite coefficient. Here ``$\dots$'' represent terms that vanish in the limit of a large region, $L\rightarrow\infty$. The subleading constant term, the \emph{topological entanglement entropy}\cite{kitaev2006,levin2006} (or just the ``topological entropy''), \emph{is universal}: it depends only on the type of topological order. More specifically, they find\cite{kitaev2006,levin2006}
\begin{equation}
\gamma = \ln \mathcal{D}
\label{eq:logD}
\end{equation}
where $\mathcal{D}$ is the so-called {\em total quantum dimension}.  For a normal state, $\mathcal{D}=1$ while for topological orders described by discrete gauge theories, $\mathcal{D}$ is the number of elements in the gauge group.  In the general case, $\mathcal{D}$ need not be an integer but in this paper we will concentrate on the simplest type of topological order, namely $\mathbb{Z}_2$ gauge theory where $\mathcal{D}=2$.  Examples of models with $\mathbb{Z}_2$ topological phases include the $\mathbb{Z}_2$ gauge theory and its extreme deconfined limit, Kitaev's toric code\cite{kitaev97}, the quantum eight vertex model\cite{ardonne04}, and the quantum dimer model on the triangular\cite{moessner2001} and other  lattices. \cite{misguich02,moessner03a} 

Eqs.~\eqref{eq:sA} and \eqref{eq:logD} were derived in Ref.~\onlinecite{kitaev2006} using topological quantum field theory, believed to capture the universal properties of topological phases, while in Ref.~\onlinecite{levin2006}, these expressions were obtained for the collection of ``string-net condensate'' wave functions, scale-invariant wave functions, believed to be ``fixed point'' models of topological order.  A common feature of both approaches is that the correlation length is (implicitly in Ref.~\onlinecite{kitaev2006}) exactly zero.  

In the present work, we study systems with finite correlation length $\xi$ to understand the effect of local correlations on the subleading part of Eq.~\eqref{eq:sA} and the issues involved with extracting the topological entropy from non-universal corrections.  Our motivation is twofold.  The first is to understand the sense in which $\gamma$ is a universal property of a topological phase.  The calculations in Refs.~\onlinecite{kitaev2006} and \onlinecite{levin2006} are based on the assumption that for systems with a finite correlation length $\xi$,  the universal behavior should apply on length scales much longer than $\xi$. In this paper we reexamine the effects of the existence of a finite correlation length in the context of specific models whose wave functions are known for all $\xi$, and verify that, indeed, this behavior is a property of the topological phase and not just of the $\xi \to 0$ limit.

Having a finite correlation length should change the value of the non-universal coefficient $\alpha$ in Eq.~\eqref{eq:sA} as well as give non-universal corrections to the constant term.  However, Refs.~\onlinecite{kitaev2006} and \onlinecite{levin2006} offered strong additional arguments for why it should still be possible to define a universal topological entropy.  In Ref.~\onlinecite{kitaev2006}, it was noted that correlation length related corrections to the subleading term would come from parts of the boundary that were not smooth (on the scale of $\xi$) such as corners.  This led them to calculate the topological entropy by partitioning the system into subregions and considering a special sum of von Neumann entropies designed to exactly cancel contributions related to the boundary and corners leaving only the universal part $\gamma$.  In Ref.~\onlinecite{levin2006}, it was noted that a defining feature of topological order is nonlocal correlation revealed by the existence of closed loop operators having nonzero expectation value.  They noted that having a finite correlation length would amount to a redefinition of the relevant loop operators, including perhaps an effective ``fattening'' of the loop.  Either way, the von Neumann entropy of an annular region would be lower than the value expected for purely local correlations and the difference approaches a universal value ($2\gamma$ for their geometry) in the large $L$ limit.  In this paper, we will show how these ideas work by explicit calculations on models with finite correlation length.

Our second motivation is that the practical importance of the topological entropy as a diagnostic for topological order is limited by the ease with which the universal part can be separated from finite size effects and other non-universal corrections due to finite correlation length, and (on a lattice) ambiguities in defining the  length of the boundary.  These latter two issues make the process more subtle than just calculating $S_A$ versus $L$ and then finding the intercept.  While such effects can be accounted for by the constructions mentioned above,\cite{kitaev2006,levin2006} there is still the problem of attaining a large enough system size that the scaling form Eq.~\eqref{eq:sA} becomes accurate.  In addition, if the calculation involves approximate methods, then a natural issue with these constructions is that $\gamma$ is a subleading term obtained by adding and subtracting large numbers.  This can make approximate calculations difficult as the errors can be comparable to the quantity of interest.  

These issues were confronted in recent numerical works aimed at verifying Eq.~\eqref{eq:logD} for the simplest examples of topological phases with finite correlation lengths.  In Refs.~\onlinecite{hamma07, furukawa07}, $\gamma=\ln 2$ was verified for the toric code\cite{kitaev97} and triangular lattice quantum dimer model at the Rokhsar-Kivelson (RK) point\cite{rokhsar88,moessner2001} respectively, by directly computing von Neumann entropies from the exact reduced density matrix. Similar calculations, performed for fractional quantum hall wave functions\cite{haque07,zozulya07}, provided values of $\gamma$ consistent with Eq.~\eqref{eq:logD}, but with errors due to finite size effects.  In this paper, through model calculations, we aim to clarify how topological information is ``stored'' in the ground state wave function and how a finite correlation length affects the entanglement.  This understanding will lead us to propose reduced quantities which should contain the same topological information as the von Neumann entropy but may be simpler to work with numerically.  While our calculations will apply for a particular class of wave functions, our conclusions should hold more generally.

In this paper, we mainly concentrate on wave functions whose amplitudes have the same form as local Boltzmann weights of related models in 2D classical statistical mechanics.\cite{rokhsar88,moessner2001,ardonne04,castelnovo05,Fendley05} In Section \ref{sec:entangle}, we will show that for such states, calculating the von Neumann entropy of region $A$ is equivalent to determining probabilities for configurations of the boundary of $A$.  (A closely related problem, the entanglement entropy at $z=2$ conformal quantum critical points in 2D,  was considered in Ref. \onlinecite{fradkin06}, which the present work generalizes.) From this perspective, in Section \ref{sec:q8v} we will consider the von Neumann entropy of the ground state of the quantum eight-vertex model near the Kitaev point\cite{ardonne04}, where the ground state wave function is equivalent to that of the toric code\cite{kitaev97}.  The Kitaev point is located inside a phase with $\mathbb{Z}_2$ topological order and at this special point, the correlation length is zero.  Away from this special point, the correlation length increases in a known manner.  In perturbation theory, we verify the assertion of Ref.~\onlinecite{kitaev2006} that finite correlation length corrections to Eq.~\eqref{eq:sA} come from appropriately defined corners of the boundary of the region under study.  In Section \ref{sec:tqdm}, we consider the ground state of the triangular lattice quantum dimer model at the RK point\cite{moessner2001,rokhsar88}.  We show that 
the topological entropy can be related to the emergence of topological \emph{winding sectors} in the calculation of the entanglement entropy.  Finally, in Section~\ref{sec:defects} we consider the effects of topological defects on the entanglement entropy of the quantum eight vertex model. We first show that mobile defects in the wave function, which drive the system into a confining phase, make the topological entropy vanish. However, in the perturbative regime where defects are slightly favored, topological entropy is shown to remain robust in low orders in the perturbation expansion series, and only corrections due to the boundary shape are allowed. Along these lines, we identify the terms in the wave function which add \emph{topological} corrections to the topological entropy, and when proliferating, should lead to a topological phase transition. Our results clearly suggest that, within the radius of convergence of perturbation theory (which is indeed finite), and for regions much larger than the correlation length $\xi$, the topological entropy $\gamma$ is a universal (and hence constant) property of the topological phase.  In Section \ref{sec:conclusions}, we conclude by discussing open problems and implications of these ideas for numerical calculations of the topological entropy. Finally, in Appendix \ref{sec:summary} we present a summary of known results of the $\mathbb{Z}_2$ gauge theory relevant to this discussion.

\section{Entanglement entropy and Measures for topological order}
\label{sec:entangle}

In this Section, we
discuss the problem of calculating the von Neumann (entanglement) entropy for a class of wave functions
that includes the ground states of various two-dimensional models of the type considered in this paper.  The key property of these wave functions is that the amplitudes of the various states are {\em local} functions of the degrees of freedom, {\it i.e.\/} the amplitudes have the form of a Gibbs weight in a suitably defined classical statistical mechanical system. Consequently, the {\em norm} of such wave functions has the form of a partition function of the related classical statistical mechanical system. The main aim of this Section is to show that the calculation of the entanglement entropy in such wave functions has an analog in the equivalent classical statistical mechanical problem. This strategy, applied to generic topological phases with finite correlation length, leads to natural generalizations of the relations recently derived in Refs.~\onlinecite{fradkin06} and \onlinecite{hamma05}.

We begin by dividing the system into simply connected subregions $A$ and $B$.  Each degree of freedom in the system belongs to either region $A$ or $B$ and the closed boundary curve separating these regions is denoted by $\Gamma$.  A convenient {\em basis} to work with is the set of product states of the form $\ket{a, b}\equiv\ket{a} \otimes \ket{b}$, where $\ket{a}$ and $\ket{b}$ denote particular configurations of the degrees of freedom in regions $A$ and $B$ respectively. This basis is complete and orthonormal.

In this basis, an arbitrary wave function will have the form $\ket{\Psi}=\frac{1}{\sqrt{Z}}\sum_{a,b}c_{a, b}\ket{a,b}$, where the $c_{a,b}$'s are complex amplitudes, and $Z=\sum_{a,b} |c_{a,b}|^2$ is the normalization of the state $\ket{\Psi}$.

The wave functions we are interested in have three additional properties:  
\begin{enumerate}
\item
The sum is over a restricted set of states that may be written in terms of the action of a set of transformations $\mathcal{G}$ on some reference state $\ket{a, b}=g\ket{a_0,b_0}$, where $g \in \mathcal{G}$.  
\item
For the states allowed in the sum, we assume the probability amplitudes are {\em local}, that is: 
\begin{eqnarray}
c_{ab}=\tilde c_a\tilde c_b
\label{factor}
\end{eqnarray}
 {\it i.e.\/}, a change of the state in region $A$ is directly connected to the corresponding change of the amplitude, irrespective of the state in region $B$.  
 \item
 We assume the set of transformations $\mathcal{G}$ form a group.   {\it i.e.\/},  The elements of $\mathcal{G}$ satisfy the four group axioms:  (a) closure.  If $g_1,g_2\in\mathcal{G}$, then the composite transformations $g_3=g_1 g_2$ and $g_3' = g_2 g_1$ 
 are also in $\mathcal{G}$ (though $g_3$ need not equal $g_3'$).  (b) associativity.  For all $g_1$, $g_2$, and $g_3$ in $\mathcal{G}$, $(g_1g_2)g_3=g_1(g_2g_3)$.  (c)  the identity transformation is in $\mathcal{G}$ and (d) If $g\in\mathcal{G}$, then so is the inverse transformation $g^{-1}$.     
 \end{enumerate}
 These properties imply that our wave functions of interest have the form: $\ket{\Psi}=\frac{1}{\sqrt{Z}}\sum_{g\in\mathcal{G}} \tilde c^{g}_{A}\tilde c^{g}_B g\ket{a_0,b_0}$,
where the reference state $\ket{a_0,b_0}$ and the group $\mathcal{G}$ encode whatever constraints the wave function respects.    The nature of the group of transformations $\mathcal{G}$ depends not only on the system but also on the representation chosen to describe the states. In the case of the systems that we discuss in this paper, the quantum dimer model, the $\mathbb{Z}_2$ gauge theory and the quantum eight vertex model, there is a local $\mathbb{Z}_2$ gauge symmetry. In all these systems there are two mutually dual representations, both of which associated with the group $\mathbb{Z}_2$, although the physical interpretation will be different. As shown in Appendix \ref{sec:summary}, in the {\em electric representation}, the Hilbert space is the space of a set of loop(``strands'') configurations. In this case, the $\mathbb{Z}_2$ group of transformations $\{ g \}$ is the Braid group with the associated Temperley-Lieb algebra, represented in the Hamiltonian by the flip (or resonance) term. In the language of the $\mathbb{Z}_2$ gauge theory, there is a one-to-one correspondence between these transformations and the set of Wilson loops of the gauge theory, $W_j=\prod_{\ell \in \Gamma_j} \sigma^z_\ell$, where $\Gamma_j$ are closed loops and $\{ \ell \}$ are the links on that loop. Conversely, in the {\em magnetic representation}, the $\mathbb{Z}_2$ group is just the group of local gauge transformations. In this picture the gauge transformations are in one-to-one correspondence with the set of dual Wilson loops $\widetilde W_k=\prod_{\widetilde \ell \in \widetilde \Gamma_k} \sigma^x_{\widetilde \ell}$, where $\widetilde \Gamma_k$ are closed loops on the dual lattice, and $\{ \widetilde \ell \}$ are links of the direct lattice intersected by $\widetilde \Gamma_k$. It is a peculiarity of systems with a $\mathbb{Z}_2$ symmetry that these dual representations are isomorphic to each other. The electric and magnetic Wilson loops obey a non-trivial algebra.\cite{freedman04}

For example, in the quantum dimer model, $\mathcal{G}$ is the set of operators whose action is to move dimers from occupied to empty links along {\em flippable} loops.  In other words, $\mathcal{G}$ is the set of operations which allow one to transform a given dimer covering of the lattice into another without ever violating the constraint that a site may have one and only one dimer.  Similarly, in the quantum eight vertex model, $\mathcal{G}$ is the set of operators whose action is to reverse the directions of arrows along flippable loops.  In other words, $\mathcal{G}$ is the set of operations which allow one to transform a given eight-vertex covering of the lattice into another without violating the constraint that a site can only have 0, 2, or 4 arrows flowing into it.

We define the reduced density matrix of region $A$ to be 
\begin{equation}
\rho_A=\Tr_{B}{\ket{\Psi}\bra{\Psi}}
\label{pure-rho}
\end{equation}
 and its von Neumann entropy is defined as:
\begin{eqnarray}
S_A=-\Tr_A\left(\rho_A\log\rho_A\right)
\label{von_neumann}
\end{eqnarray}
A useful way to calculate $S_A$ is through the so-called ``replica trick''\cite{holzhey94,calabrese04}:
\begin{eqnarray}
S_A=-\lim_{n\rw1}\frac{\pd}{\pd n}\bigg(\Tr_A \rho_{A}^n\bigg)\equiv \lim_{n \to 1}S_{A,n}
\label{entrlim}
\end{eqnarray}
where $\rho_A^n$ is the $n$-th power of the reduced density matrix $\rho_A$. 

It is an implicit assumption of this ``replica'' approach that there is a well defined and unique {\em analytic continuation} of the generalized ``entropies'' $S_{A,n}$ defined for $n$ in the set of positive integers to a region of the complex plane which includes $n=1$. As we shall see, this assumption holds for the cases we are interested in. (We note, however, that there are systems, studied in the recent literature,\cite{casini05,cardy07} for which the existence of the $n \to 1$ limit is quite non trivial.)

For a general wave function, $\Tr_A\rho_{A}^n$ is given by:
\begin{eqnarray}
&&Z^n\Tr_A \rho_{A}^n =\nonumber \\
&&\sum_{a_1,b_1, \ldots, a_n, b_n} c_{a_1,b_1}c^{*}_{a_2,b_1}c_{a_2,b_2}c^{*}_{a_3,b_2}\cdots c_{a_n,b_n}c^{*}_{a_1,b_n}
\nonumber \\
&&
\label{eq:matchf}
\end{eqnarray}
For a constrained wave function, $c^{*}_{a_2,b_1}$ will be zero unless $\ket{a_2,b_1}$ is one of the allowed configurations in the ground state sum and so on.  If in addition, we assume the condition \eqref{factor}, then the expression may be written:
\begin{eqnarray}
\Tr_A\rho_{A}^n&=&\frac{1}{Z^n}\sum_{a_1,b_1\cdots a_n,b_n}^{\; \prime} |\tilde c_{a_1}|^2|\tilde c_{b_1}|^2|\tilde c_{a_2}|^2\cdots|\tilde c_{a_n}|^2|\tilde c_{b_n}|^2 \nonumber\\ 
\label{eq:trra1}
\end{eqnarray}
where the prime denotes that the sum over $b_1$ is restricted to states where both $\ket{a_1,b_1}$ and $\ket{a_2,b_1}$ are allowed configurations and likewise for $b_2$ and so on.  
If we view the labels $1,2...,n$ as denoting $n$ different copies of our two-dimensional system, then the sum in Eq.~\eqref{eq:matchf} can be interpreted as a partition function for $n$ systems where region $A$ of copy $k+1$ is constrained to be consistent with region $B$ of copy $k$ and so on.   

An additional simplification arises if the allowed states in the wave function are related by the group property.  In this case, we have ``transitivity'', {\it i.e.\/} if $\ket{a_1,b_1}$, $\ket{a_2,b_1}$, and $\ket{a_2,b_2}$  are allowed configurations, then so is $\ket{a_1,b_2}$.  The implication is that the sum in Eq.~\eqref{eq:trra1} is the partition function for $n$ copies that are constrained such that region $A$ of any copy is consistent with region $B$ of any other copy.  We may write this as:
\begin{eqnarray}
\Tr_A\rho_{A}^n = \frac{Z[n]}{Z[1]^n}
\label{eq:trra}
\end{eqnarray}
where $Z[n]$ is the partition function of $n$ copies constrained to agree with each other on the boundary $\Gamma$ (in the sense discussed above) but are otherwise independent and $Z[1]=Z$ is the partition function of a single copy. This result was derived in Ref.~\onlinecite{fradkin06} by a different line of arguments.

The most trivial way that the $n$ copies agree on the boundary $\Gamma$ is if the state of the boundary is the same in all of them.  However, this is not the most general way of agreement.  
In general, as we noted above, the operators $g$ can be regarded as either products of Wilson loop operators $W_j$, or of the dual Wilson loops, $\widetilde W_k$, of the corresponding gauge theory. Such a Wilson loop operator might involve degrees of freedom in region A ($W^{(A)}$), in region B ($W^{(B)}$), or both $(W^{(AB)})$. What is clear now is that the agreement of all $n$ copies persists if no $W^{(AB)}$ operators transform any of the copies. This means that the degrees of freedom on the boundary $\Gamma$ might change through application of $W^{(A)}$ or $W^{(B)}$ and the agreement of the $n$ copies is not affected. The loop operators $W^{(AB)}$ naturally form equivalence classes, each defined by the set of loops $W^{(AB)}$ modulo contractible loops defined in either region $A$ or $B$ but not in both. Each class is labelled by an irreducible loop $W^{(AB)}$.  However, if any of the possible irreducible operators $W^{(AB)}$  is applied on any of the copies, then the agreement is violated and it is restored only if the same $W^{(AB)}$ or any of its reducibles is applied on all the other copies.

As a concrete example, consider the quantum dimer model where the boundary loop $\Gamma$ is drawn on the direct lattice (see Fig.~\ref{tdbound}).  We consider the lattice {\em links} inside and along the boundary as belonging to region $A$ and the links outside as part of region $B$.  As will be shown in Section \ref{sec:tqdm}, we can classify dimer configurations by looking at the lattice {\em sites} which lie along $\Gamma$ and for each of these sites, specifying whether its dimer belongs to region $A$ or $B$.  Two dimer coverings that agree along $\Gamma$ {\em in this manner} \footnote{Note that this is not the same as requiring the two dimer coverings to have the {\em same} dimer arrangement on the boundary.} may be said to ``agree on the boundary'' in the sense discussed above:  if attempt to make a new dimer covering by combining the dimer pattern of region $A$ of the first copy with that of region $B$ of the second copy, the new configuration will automatically satisfy the hardcore dimer constraint on the boundary.  It is easy to see that flipping the dimers from occupied to empty links along loops which live solely on links of region $A$ or $B$, i.e. acting on the configuration with operators $W^{(A)}$ or $W^{(B)}$ in the above paragraph, will not change the state of the boundary as just defined.  However, performing this operation along a loop which intersects the boundary, i.e. acting on the configuration with operators $W^{(AB)}$ in the above paragraph, will change the state of the boundary.   

Returning to the general case, it is useful to classify configurations based on the states of their boundaries.  This procedure leads to the following identification:
\begin{eqnarray}
Z[n]=\sum_{g\in \{G^{(AB)}\}}\left(\tilde Z[1]^{g}_\Gamma\right)^n
\label{zn}
\end{eqnarray}
To explain the notation, we choose a reference configuration which also defines a reference configuration of the boundary.  As mentioned above, we can then group the various transformations $G^{(AB)}$ which will change the boundary configuration into equivalence classes based on the final state of the boundary.  We denote the collection of equivalence classes by $\{ G^{(AB)} \}$ and within each class, we may select some $g$ as a representative element.  The sum in Eq.~\ref{zn} is over the set of these representative $g$'s, which is an irreducible set.  
The partition function $\tilde Z[1]^{g}_\Gamma$ is defined as 
\begin{eqnarray}
&& \tilde Z[1]^{g}_\Gamma = Z^{g}_A Z^{g}_B=\left(\sum_{a}|\tilde c_{a}|^2\right)\left(\sum_{b}|\tilde c_{b}|^2\right) 
\nonumber \\
&&
\end{eqnarray}
where $a,b$ run along all possible configurations of any of the copies which are produced by applying products of $W^{(A)}$ or $W^{(B)}$ operators, respectively, on a reference configuration.  In other words,
$\tilde Z[1]^{g}_\Gamma$ is the partition function of a {\em single} copy of the system (note $Z[1]$ as opposed to $Z[n]$) where the configuration of the boundary $\Gamma$ is now {\em fixed} to the configuration obtained when transformation $g$ acts on the reference state.

In this way, if we use \eqref{entrlim}, \eqref{eq:trra} and \eqref{zn}, we have:
\begin{eqnarray}
S_A=-\sum_{g\in \{G^{(AB)}\}} p^{g}_\Gamma\log(p_{\Gamma}^g)
\label{enen}
\end{eqnarray}
where $p^{g}_\Gamma=\tilde Z[1]^{g}_\Gamma/Z[1]$.

From another point of view, given that  the sum in Eq. \eqref{enen} runs over all possible irreducible (either gauge or Temperley-Lieb, depending on the case) transformations $g\in \{ G^{(AB)}\}$ of the boundary configuration, involving degrees of freedom in \emph{both} regions A and B, we have:
 \begin{eqnarray}
 S_A=-\left<\log\left(\frac{Z_{A}^gZ_{B}^g}{Z[1]}\right)\right>_{g}
 \label{engauge}
 \end{eqnarray}
 The form of Eq.~\ref{engauge} is quite instructive, showing that the entanglement entropy is an average quantity over gauge transformations, as one should expect, since the boundary has no special properties.

In a topologically ordered state (non-critical) which is defined by a ground state wave function which satisfies the property in Eq.\eqref{factor}, the calculation of the entanglement entropy leads to the general result\cite{kitaev2006,levin2006}:
 \begin{eqnarray}
 S_A=a(\xi)L+b(\xi)-\gamma+O(1/L)
 \label{sAgeneral}
 \end{eqnarray}
 where $a(\xi),b(\xi)$ are correlation length dependent constants. By the argument of Kitaev and Preskill\cite{kitaev2006} and Levin and Wen\cite{levin2006}, the constant term $\gamma$ is expected to be a topological invariant and as such it should remain constant throughout the topological phase. Its value is known in the topological limit,where the correlation length vanishes, to be\cite{kitaev2006,levin2006}  $\gamma=\ln\sqrt{\sum_{i}d_{i}^2}$, where $d_{i}$ is the quantum dimension of particles of type $i$ and the sum is over all non-trivial superselection sectors of the medium. 
 
  On the other hand, the average over all possible transformations in Eq. \eqref{engauge} shows that ``measuring'' the entanglement entropy of a region of the medium does not endow its boundary with any special property that would make it different from any other place in the system. Motivated by these observations, we note that since $\gamma$ is a topological invariant, its value should in fact be the same for all members of the ensemble of transformations implied in Eq.\eqref{engauge}. Thus, it should be possible to obtain this quantity by ``fixing completely the gauge'' on the boundary between regions $A$ and $B$ which should not affect the topological properties of the state. In this way, we define the following reduced quantity:
 \begin{eqnarray} 
 S^{(s)}_A=-\log\left(\frac{Z_{A}^{g_0}Z_{B}^{g_0}}{Z[1]}\right)
 \label{suben}
 \end{eqnarray}
  where $g_0$ is a single choice of all possible $G^{(AB)}$s.
 We conjecture that $S^{(s)}_A$ contains the same topological information as $S_A$. In other words:
 \begin{eqnarray}
 S^{(s)}_A=\tilde a(\xi)L + \tilde b(\xi)-\gamma + \tilde O(1/L)
 \label{conjform}
 \end{eqnarray}
The calculation of this quantity involves an effective splitting of the system, similar to the procedure that was used in Ref. \onlinecite{fendley06b}. 
On the other hand, the fact that $S^{(s)}_A$ has the form of Eq.\eqref{conjform}  can be trivially proven for topological systems with zero correlation length, such as the ones in Refs.~\onlinecite{hamma05,kitaev2006,levin2006}. Moreover, we verify in the following sections that such a form still holds for systems with a finite-correlation length, such as the quantum eight-vertex model of Ref. \onlinecite{ardonne04} in its topological phase, in the deconfined phase of the $\mathbb{Z}_2$ gauge theory, and in topologically ordered ground states of quantum dimer models at the RK points\cite{rokhsar88,moessner2001}.
 
From a numerical point of view, the advantage of this reduced quantity is that the whole set of matrix elements of the density matrix of the full system is not necessary. Instead, one needs the density matrix of a specially \emph{local} wave function, making the task of extracting $\gamma$ from a wave function much simpler. Moreover, the constructions by Levin-Wen and Kitaev-Preskill can be used for $S^{(s)}_A$ as well, but now the number of additions-subtractions of large numbers that are needed is minimized.

The use of the representation of entanglement entropy of Eq. \eqref{engauge} and the consequent reduction of Eq. \eqref{suben} is formally possible only when Eq. \eqref{factor} holds. Eq. \eqref{factor} expresses the requirement that regions $A$ and $B$ are interacting ultra-locally, only through the degrees of freedom of the boundary $\Gamma$. However, more generally, one should consider ground state wave functions where the amplitudes $c_{a,b}$ are defined as $c_{a,b}=\tilde c_a \tilde c_b \tilde c^{\rm int}_{ab}(d)$, where $\tilde c^{\rm int}_{ab}(d)$ depends on the relative configurations of regions $A$, $B$ at a distance $d$ from the boundary $\Gamma$. The distance $d$ characterizes microscopic properties of the ground state wave function and typically $d\ge a$ where $a$ is the lattice spacing. The distance $d$ should not be confused with the correlation length of the system $\xi$ which could become infinite. Instead, it should be viewed as a consequence of local short-range (of the order of $d$) interacting terms of the Hamiltonian, whose ground state we consider. To make Eqns. \eqref{engauge} and \eqref{suben} applicable in these cases, the definition of the reduced density matrix should be modified. If $\ket{\Gamma^i}$ represents a generic configurational state of the {\em region} $\Gamma$ near the boundary (at distance d) in region $A$, then we define:
\begin{eqnarray}
\rho^{(i)}_{\tilde A}=\bra{\Gamma_i}\Tr_{B}\rho\ket{\Gamma_i}
\end{eqnarray}
In this way, all degrees of freedom in region $B$ are integrated out and the degrees of freedom in region $\Gamma$ are fixed. $\rho^{(i)}_{\tilde A}$ is a reduced density matrix of region $\tilde A$. Now, the generalized entanglement entropy of region $\tilde A$ using $\rho^{(i)}_{\tilde A}$ as the reduced density matrix is :
\begin{eqnarray}
\tilde S_{\tilde A}=-\sum_{i}\Tr_{\tilde A}\left[\rho^{(i)}_{\tilde A}\ln\rho^{(i)}_{\tilde A}\right]
\label{enen2}
\end{eqnarray}
where the sum runs along all possible configurations of region $\Gamma$. Now, Eqns. \eqref{eq:trra} and \eqref{zn} can be applied without any change, because the property in Eq. \eqref{factor} holds for regions $\tilde A, B$ and thus, a partition function representation is possible. Following the same type of reasoning, we end up to the following relation:
\begin{eqnarray}
\tilde S_{\tilde A} = -\left< \ln\frac{Z^{i(\Gamma)}_{\tilde A}Z^{i(\Gamma)}_{B}}{Z[1]}\right>_{i(\Gamma)}
\end{eqnarray}
where now $i(\Gamma)$ labels configurational states of region $\Gamma$. Following the same type of reasoning as above, we argue that, if $i_0(\Gamma)$ labels a specific configuration of region $\Gamma$, then the reduced quantity:
\begin{eqnarray}
\tilde S^{(s)}_{\tilde A} = -  \ln\frac{Z^{i_0(\Gamma)}_{\tilde A}Z^{i_0(\Gamma)}_{B}}{Z[1]}
\end{eqnarray}
contains the same topological information as $S_{A}$.

Even though such a generalized quantity, where the boundary is effectively ``fattened'', should be used for the case of general wave functions, in the case of the quantum eight-vertex model or quantum dimer models at the RK point this is not necessary.

\section{ Quantum Eight-Vertex model near the Kitaev point} 
\label{sec:q8v}

In this Section we compute the entanglement entropy of the ground state wave function of the quantum eight-vertex model of Ref. \onlinecite{ardonne04} in its quantum disordered, topological, phase. We will follow the methods discussed in the previous Section.

\begin{figure}[hbt]
\subfigure[]{\includegraphics[width=0.4\textwidth]{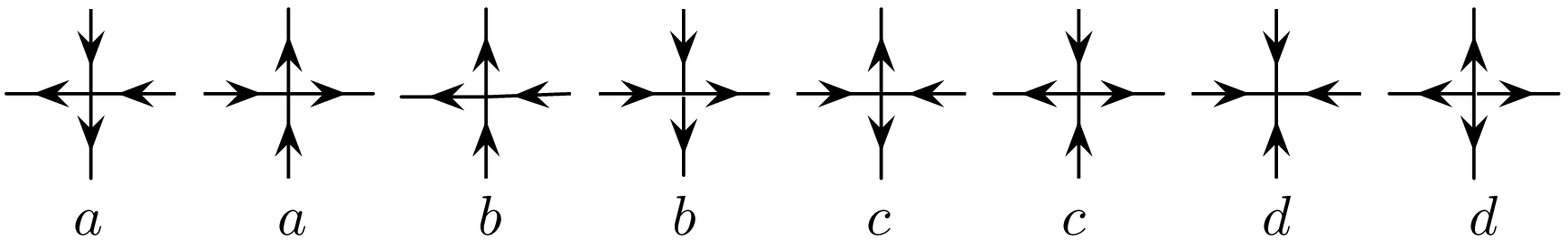}\label{8vs}}
\subfigure[]{\includegraphics[width=0.2\textwidth]{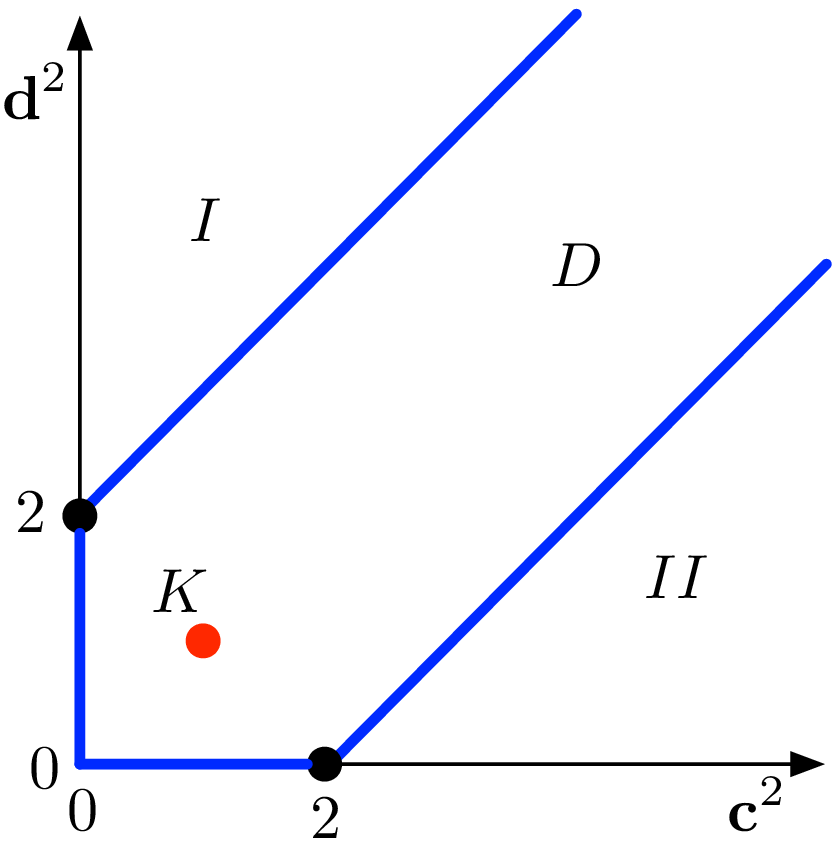}\label{8vpd}}
\caption{(a) Allowed vertices in the classical eight-vertex model. (b) The phase diagram of the classical eight-vertex model. The phases I and II represent ``antiferroelectric '' phases dominated by $d$ and $c$ vertices respectively. The middle D phase is disordered.  The point $K$ is where $c=d=1$.  In the quantum version of the model, the ground state wave function at this ``Kitaev point'' is the same as the ground state of Kitaev's toric code\cite{kitaev97}.}
\end{figure}

The quantum eight-vertex model of Ref. \onlinecite{ardonne04} is a a two-dimensional quantum generalization of the classical two-dimensional eight vertex model\cite{baxterbook}, which is integrable.  In the classical model, the degrees of freedom are arrows that live on the links of a square lattice; arrows on horizontal (vertical) links point left or right (up or down).  In addition, there is a local constraint that each vertex may have only an even number of ingoing (or outgoing) arrows. 
This results in eight types of allowed vertices. (cf. Fig.\ref{8vs})
The partition function of the classical model is:
\begin{eqnarray} 
Z_{8V}=\sum_{\cC}a^{2n_a(\cC)}b^{2n_b(\cC)}c^{2n_c(\cC)}d^{2n_d(\cC)}
\label{zcl8v}
\end{eqnarray}
Due to global arrow neutrality (the ingoing equal the outgoing arrows), the weights $a,b$ are redundant, and can be set to $1$. The phase diagram of this model in terms of the weights $c$ and $d$ is shown in Fig.\ref{8vpd}. There are two ordered phases for $c^2>d^2+2$ and $d^2>c^2+2$ separated from a disordered phase by lines of continuously varying exponents ($c^2=d^2+2$ and $d^2=c^2+2$). Moreover, for $c=0,d^2<2$ and $d=0,c^2<2$, there are additional lines of fixed points. The region of interest for our calculation is the vicinity of the point $c=d=1$. 

The quantum generalization of this model begins by placing Pauli matrices $\si^{a}_k$ on the links of the square lattice where $k$ labels the link and $a=1,2,3$ labels the Pauli matrix.  We work in the representation where $\si^1$ is diagonal and up spins correspond to arrows pointing up or right.
In this language, the eight-vertex constraint corresponds to:
\begin{eqnarray} 
\si_{{\mathbf r-\hat x/2}}^{1}\si_{{\mathbf r + \hat x/2}}^{1}\si_{{\mathbf r + \hat y/2}}^{1}\si_{{\mathbf r - \hat y/2}}^{1}\ket{G} =\ket{G}
\end{eqnarray}
where $\hat x$ and $\hat y$ are lattice unit vectors.  This expression holds for any vertex ${\mathbf r}$ and $\ket{G}$ is a state within the manifold spanned by the eight vertex configurations.  We then consider the following Hamiltonian:
\begin{eqnarray}
\hat H_{K}=-t\sum_{\Box}\si_{{\mathbf r+\hat x/2}}^{3}\si_{{\mathbf r + \hat x+\hat y/2}}^{3}\si_{{\mathbf r +\hat x/2+\hat y }}^{3}\si_{{\mathbf r+\hat y/2 }}^{3}
\label{hk}
\end{eqnarray}
where the sum is over plaquettes.  The individual terms are operators that flip the spins (or equivalently the arrows) around a plaquette.  These flip operators commute with the constraint so it is possible to diagonalize both simultaneously.  As shown in Refs.~\onlinecite{kitaev97} and \onlinecite{ardonne04}, for $c=d=1$  
this results in a ground-state wave function given by:
\begin{eqnarray}
\ket{G_K}= \frac{1}{\sqrt{Z_{8V}}}\sum_{\{\cC\}}\ket{\cC}
\end{eqnarray}
where $\cC$ corresponds to an eight-vertex configuration of the square lattice. This wave function is exactly  the ground-state wave function of Kitaev's toric code. The norm of this wave function corresponds to the partition function of the eight-vertex model at $c=d=1$, which we label the Kitaev point (K in Fig.~\ref{8vpd}). Moreover, in Ref.\onlinecite{ardonne04}, it was shown that the Hamiltonian $H_K$ can be \emph{locally} deformed in such a way that the ground state wave function now becomes:
\begin{eqnarray}
\ket{G_{8V}} = \frac{1}{\sqrt{Z_{8V}}}\sum_{\{\cC\}}c^{\hat n_c(\cC)}d^{\hat n_d(\cC)} \ket{\cC}
\label{g8v}
\end{eqnarray}
where now $c,d$ are free parameters of the Hamiltonian and:
\begin{eqnarray}
\hat n_c(\cC)\equiv \frac{1}{16}\sum_{\mathbf r}\bigg[\si_{{\mathbf r+\hat x/2}}^{1}-\si_{{\mathbf r-\hat x/2}}^{1}-\si_{{\mathbf r+\hat y/2}}^{1}+\si_{{\mathbf r-\hat y/2}}^{1}\bigg]^2 \label{nc}\nonumber\\
\end{eqnarray}
\begin{eqnarray}
\hat n_d(\cC)\equiv \frac{1}{16}\sum_{\mathbf r}\bigg[\si_{{\mathbf r+\hat x/2}}^{1}-\si_{{\mathbf r-\hat x/2}}^{1}+\si_{{\mathbf r+\hat y/2}}^{1}-\si_{{\mathbf r-\hat y/2}}^{1}\bigg]^2\label{nd}\nonumber\\
\end{eqnarray} 
The norm of this state, $\braa{G_{8V}}\ket{G_{8V}}$, is just the partition function of the classical two-dimensional eight-vertex model, Eq. \eqref{zcl8v} and thus, the ground-state phase diagram is identical to the classical one. The main difference  is that what in the classical system is the disordered phase, in the quantum system is the topological phase.  

We also may rewrite the ground-state wave function \eqref{g8v} in a more helpful and interesting way:
\begin{eqnarray}
\ket{G_{8V}} = \frac{1}{\sqrt{Z_{8V}}}\sum_{\{g\}}c^{\hat n_c(g)}d^{\hat n_d(g)} g\ket{0}
\label{g8v2}
\end{eqnarray}
and its norm:
\begin{eqnarray} 
Z_{8V}=\sum_{\{g\}}c^{2\hat n_c(g)}d^{2\hat n_d(g)}
\end{eqnarray}
where $\ket{0}$ represents the ``vacuum'' state, where $\si_{\mathbf r}^1\ket{0}=\ket{0}$ for any ${\mathbf r}$ and $\ket{g}\equiv g\ket{0}$ corresponds to the repeated actions of the flip operator on the vacuum state. More specifically, we may think of $g$ as a product of  single plaquette-flip operators $\prod_{\Box}\si^{3}_{\mathbf r}$ and then, $\ket{g}$ differs from $\ket{0}$ only in that some plaquettes have been flipped. It is important to notice that the single plaquette-flip operator can be applied to all but one plaquette, because the product of all plaquette-flip operators on a closed surface is equal to the identity operator. As we noted above, and explained in Appendix \ref{sec:summary}, the flip operator is a representation of the Temperley-Lieb generator acting on the loops  (or strands) of the eight-vertex model.

\begin{figure}[h]
\subfigure[]{\includegraphics[width=0.2\textwidth]{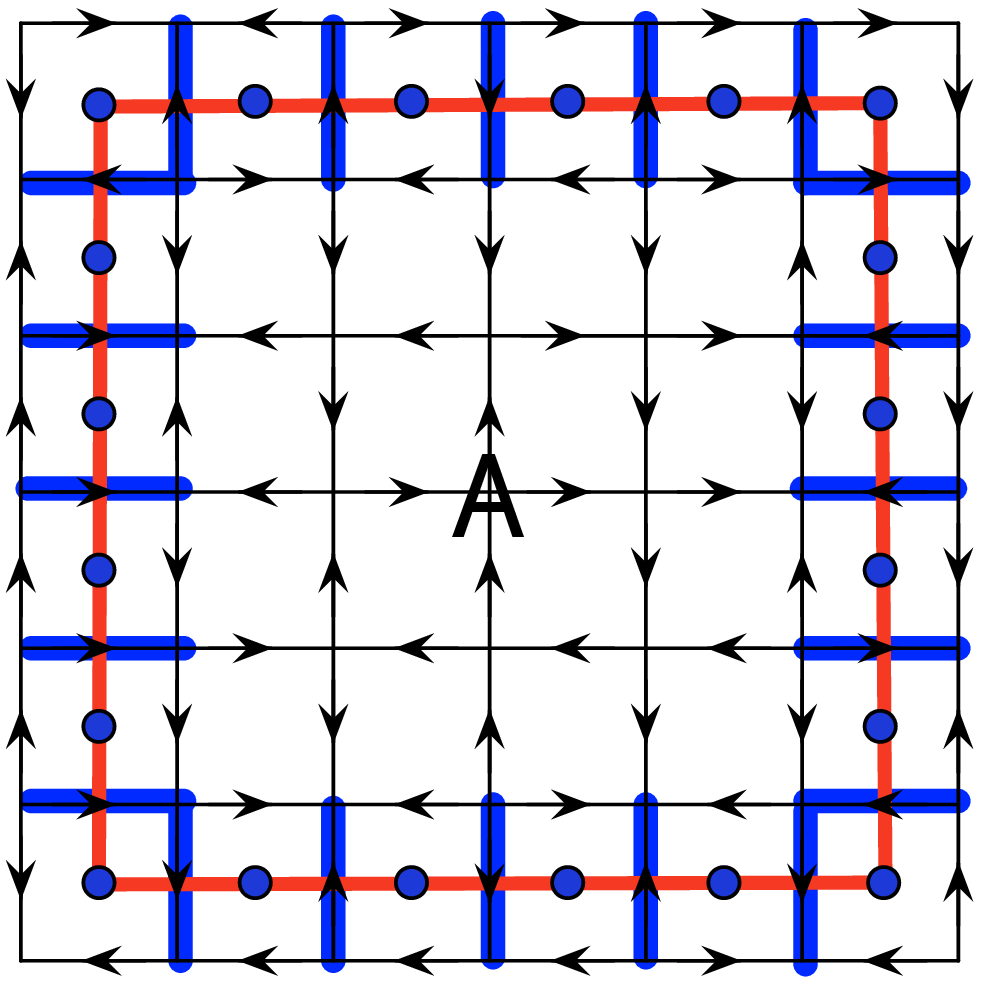}\label{boundary}}
\subfigure[]{\includegraphics[width=0.18\textwidth]{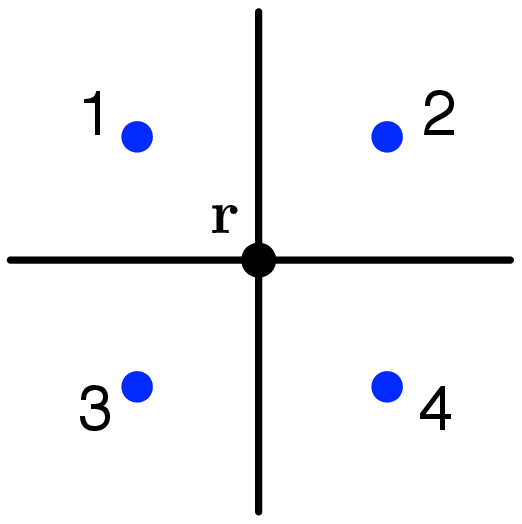}\label{directdual}}
\caption{(a) A pictorial representation of the boundary line $\Gamma$ and the dual lattice variables which have to remain frozen in the calculation of the entanglement entropy. (b) The dual lattice variables with respect to a lattice site of the direct lattice.}
\end{figure}

 Now, we can label each plaquette with an Ising variable $\tau^{g}_{ \tilde {\mathbf r}}$, living at the center of each elementary plaquette, which will indicate in an eight-vertex configuration whether the plaquette-flip operator is applied or not. In the vacuum state $\ket{0}$, $\tau^{0}_{\tilde {\mathbf r}}=-1$ for any plaquette. In this representation, we have:
 \begin{eqnarray}
 \si^{1 g}_{{\mathbf r}+\hat y/2}&\equiv& \tau^{g}_{\tilde{\mathbf r}-\hat x}\tau^{g}_{\tilde{\mathbf r}}\equiv\tau^{g}_{{\bf r}, 1}\tau^{g}_{{\bf r}, 2}\label{sitau1}\\
 \si^{1g}_{{\mathbf r}+\hat x/2}&\equiv& \tau^{g}_{\tilde{\mathbf r}-\hat y}\tau^{g}_{\tilde{\mathbf r}}\equiv\tau^{g}_{{\bf r}, 3}\tau^{g}_{{\bf r}, 4}\label{sitau2}
 \end{eqnarray} 
 where the indices $1,2,3,4$ are defined in Fig.\ref{directdual}. In this representation, the density variables $n_c$,$n_d$ \eqref{nc},\eqref{nd}, using the definitions \eqref{sitau1}, \eqref{sitau2} can be expressed as follows:
\begin{eqnarray}
n_c(g) &= &\frac{1}{4}\sum_{\bf r}\big(1+\tau^{g}_{{\bf r},2}\tau^{g}_{{\bf r},3}-\tau^{g}_{{\bf r},1}\tau^{g}_{{\bf r},4}-\tau^{g}_{{\bf r},1}\tau^{g}_{{\bf r},2}\tau^{g}_{{\bf r},3}\tau^{g}_{{\bf r},4}\big)
\nonumber \\
n_d(g) &=& \frac{1}{4}\sum_{\bf r}\big(1-\tau^{g}_{{\bf r},2}\tau^{g}_{{\bf r},3}+\tau^{g}_{{\bf r},1}\tau^{g}_{{\bf r},4}-\tau^{g}_{{\bf r},1}\tau^{g}_{{\bf r},2}\tau^{g}_{{\bf r},3}\tau^{g}_{{\bf r},4}\big)
\nonumber \\
&&
\end{eqnarray}

These expressions will simplify the calculation of the entanglement entropy.

\subsection{Entanglement entropy near the Kitaev point}
\label{sec:entanglement-8vertex}

To compute the entanglement entropy in the disordered phase of the eight-vertex wave function we begin by dividing the eight-vertex lattice into two regions $A$ and $B$, separated by a single closed boundary line $\Gamma$ on the dual lattice (Fig.~\ref{boundary}).  We will follow the approach of Section \ref{sec:entangle}. The degrees of freedom are arrows that live on the links in region $A$ or $B$.  As shown in Fig.~$\ref{boundary}$, region $A$ contains all the degrees of freedom (links) that define the boundary, but the $\tau$ variables living at the dual sites around the boundary, pointed out in the figure, have to be fixed in the calculation of $p_{\Gamma}^g$, according to the definition we gave for $p_{\Gamma}^g$.  An important feature of this choice of the boundary is that the species ({\it i.e.\/} whether or not it is a $c$ vertex) of a vertex in region $A$ is determined solely by the links in region $A$ and similarly for the vertices of region $B$ (the eight vertex constraint means that knowing three links automatically determines the fourth).  Therefore, the probability weights of the wave function have the property \eqref{factor} which was needed in order to think of the calculation in terms of boundary probabilities.  

At the  $K$ point ($c=d=1$), for any choice of the $\tau$ variables on the boundary, the probability $p_{\Gamma}^g$ is:
\begin{eqnarray}
p_{\Gamma}^g = \frac{2^{N_s-N_\Gamma}}{2^{N_s-1}}=\left(\frac{1}{2}\right)^{N_\Gamma-1}
\label{kitaevpg}
\end{eqnarray} 
where $N_s$ is the total number of sites in the (direct or dual) lattice; $N_\Gamma$ is the number of direct lattice links or dual lattice sites on the boundary, and the $-1$ originates in the global constraint:
\begin{eqnarray} 
\prod_{\mathbf r} \si_{\mathbf r}^3 = 1
\end{eqnarray}

This result leads to the well known result for the entanglement entropy of the Kitaev's wave function:
\begin{eqnarray}
S_A=-\sum_{g}p_{\Gamma}^g\ln{p_{\Gamma}^g}=(\ln2)L-\ln2
\label{kitaevee}
\end{eqnarray}

As soon as $c\neq 1$ and $d\neq1$, the correlation length $\xi$ of the ground-state wave function becomes non-zero. In particular, as the critical line is approached on the left or right, the correlation length is known\cite{baxter71} to diverge as $\xi\sim||c-d|-2|^{-\pi/\big|4(\tan^{-1}(\sqrt{cd})\big|}$. We are interested in the effects that the correlation length has on the entanglement entropy. Now, each  transformation $g$ has a ``Gibbs weight'' which makes it more or less favorable than others. We argue that any  other way of introducing a correlation length is going to cause qualitatively similar effects to the ones that we describe here. The reason is that a correlation length is expressed generally by the fact that some eigenstates in some basis are more favored than others. We can study these effects in what in terms of the related classical eight-vertex model amounts to the (rapidly convergent) high-temperature-type expansion (we will consider the case $c\neq1,d=1$). 

We have:
\begin{eqnarray} 
c^{2 n_c(g)}&=&c^{\frac{N_s}{2}}\left[\cosh(\frac{1}{4}\ln c^2)\right]^{3N_s}\nonumber \\
&& \!\!\!\!\! \!\!\!\!\! \times \prod_{\bf r}\left(u+v\tau^{g}_{{\bf r}, 2}\tau^{g}_{{\bf r}, 3}-v\tau^{g}_{{\bf r}, 1}\tau^{g}_{{\bf r}, 4}
-v\tau^{g}_{{\bf r}, 1}\tau^{g}_{{\bf r}, 2}\tau^{g}_{{\bf r}, 3}\tau^{g}_{{\bf r}, 4}\right)  \nonumber \\
&&
\end{eqnarray}
where $u=1+\left[\tanh\frac{1}{4}\ln(c^2)\right]^3$ and $v=\tanh\frac{1}{4}\ln(c^2)+\left[\tanh\frac{1}{4}\ln(c^2)\right]^2$.
The probability $p_{\Gamma}^g$ is defined as $p^{g}_\Gamma=Z_{8V}^{\Gamma_g}/Z_{8V}$ where the index $\Gamma_g$ indicates that the boundary degrees of freedom shown in Fig.~\ref{boundary} are fixed. 

In the expansion of $Z_{8V}$ and $Z_{8V}^{\Gamma_g}$, there are two ways of taking non-trivial terms. Firstly, similarly to a high-temperature expansion of an Ising model's partition function, we have terms which are products of squared Ising variables $\tau$. Moreover, in the expansion of $Z_{8V}^{\Gamma_g}$, terms which are products of the Ising variables on the boundary $\Gamma$ (cf. Fig.~\ref{boundary}), are also non-zero, because the Ising variables on the boundary are fixed to a specific value (according to the form of the transformation $g$, which in this case corresponds to the possible different values of the Ising variables $\{\tau^{\Gamma}\}$).  These terms are classified according to the relative orientation of the appearing $\tau^{\Gamma_g}$ variables and contribute differently when the $\tau$ variables are located near one of the four corners of the boundary (cf. Fig.~\ref{boundary}).

We first expand the denominator of the partition function, keeping the most relevant term. The second most relevant term is of order $O(u^{N_s-4}v^4)\sim O([\ln c^2]^4)$:
\begin{eqnarray}
Z_{8V}&=&\nonumber \\
&&\!\!\!\!\!\! \!\!\!\!\!\! 2^{N_s-1}c^{\frac{N_s}{2}}\left[\cosh(\frac{1}{4}\ln c^2)\right]^{3N_s}\!\!\!\!\!
u^{N_s}\left(1+O\left(\left(\frac{v}{u}\right)^4\right)\right)
\nonumber \\
&&
\end{eqnarray}
In expanding the numerator, we have to keep in mind that the $\tau$ variables living on the dual sites next to the boundary, shown in Fig. \ref{boundary}, are deliberately fixed. So, expanding, we have:
\begin{eqnarray}
&&Z_{8V}^{\Gamma_g}=2^{N_s-N_\Gamma}c^{\frac{N_s}{2}}\left[\cosh(\frac{1}{4}\ln c^2)\right]^{3N_s}u^{N_s}
\nonumber\\
&&\times \Bigg\{1
+\sum_{\bf r}\Bigg[\frac{v}{u}\left({\parbox[c]{0.2in}{\includegraphics[width=0.2in]{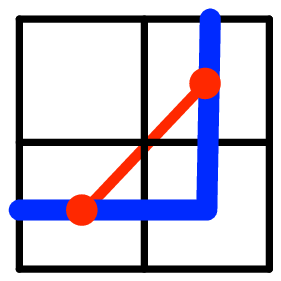}}}_{\bf r}-{\parbox[c]{0.2in}{\includegraphics[width=0.2in]{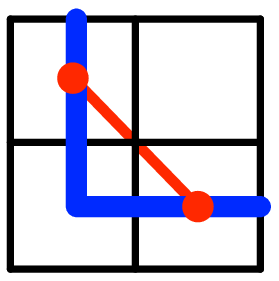}}}_{\bf r}\right)
\nonumber \\
&&+\left(\frac{v}{u}\right)^2\Bigg({\parbox[c]{0.22in}{\includegraphics[width=0.22in]{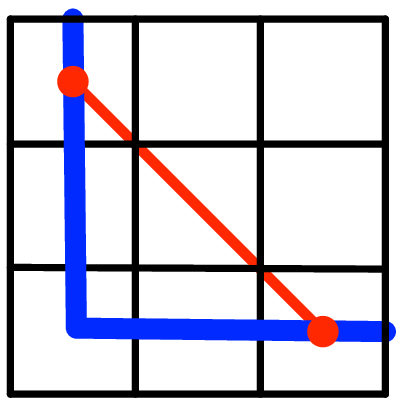}}}_{\bf r}+ {\parbox[c]{0.22in}{\includegraphics[width=0.22in]{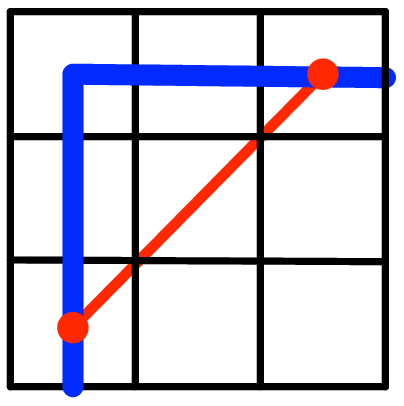}}}_{\bf r}
+ {\parbox[c]{0.22in}{\includegraphics[width=0.22in]{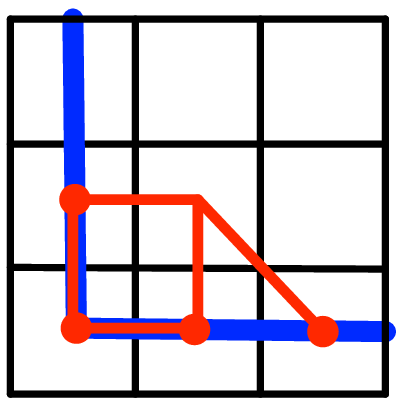}}}_{\bf r}- {\parbox[c]{0.22in}{\includegraphics[width=0.22in]{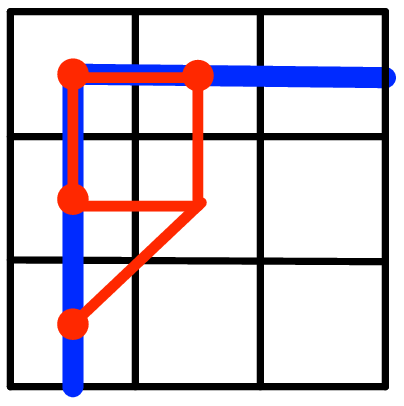}}}_{\bf r}
{-\parbox[c]{0.22in}{\includegraphics[width=0.22in]{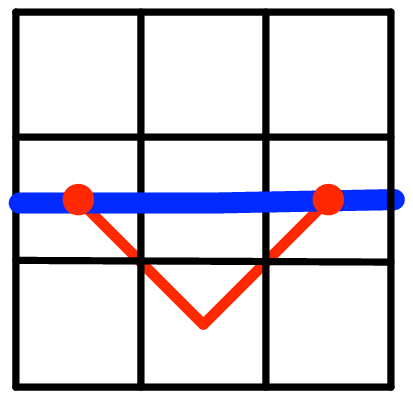}}}_{\bf r}-{\parbox[c]{0.22in}{\includegraphics[width=0.22in]{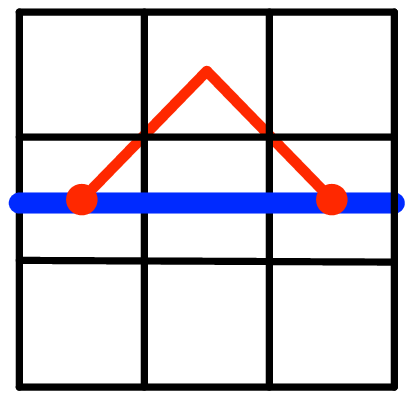}}}_{\bf r}
\nonumber\\
&&
+\frac{1}{2}\sum_{{\bf r'}\neq {\bf r}}\left({\parbox[c]{0.2in}{\includegraphics[width=0.2in]{Z8vg1.eps}}}_{\bf r}{\parbox[c]{0.2in}{\includegraphics[width=0.2in]{Z8vg1.eps}}}_{\bf r'}
-{\parbox[c]{0.2in}{\includegraphics[width=0.2in]{Z8vg1.eps}}}_{\bf r}{\parbox[c]{0.2in}{\includegraphics[width=0.2in]{Z8vg12.eps}}}_{\bf r'}+{\parbox[c]{0.2in}{\includegraphics[width=0.2in]{Z8vg12.eps}}}_{\bf r}{\parbox[c]{0.2in}{\includegraphics[width=0.2in]{Z8vg12.eps}}}_{\bf r'}\right)\Bigg)\Bigg]
\nonumber \\
&&+
O\left(\left(\frac{v}{u}\right)^3\right)\Bigg\}\;\;\;
\nonumber \\
&&
\label{pfb}
\end{eqnarray}
where ${\parbox[c]{0.22in}{\includegraphics[width=0.22in]{Z8vg1.eps}}}_{\bf r}\equiv \tau\tau$ and the variables $\tau$ are defined on the sites of the dual lattice that are pointed out around the site of the direct lattice located at $\bf r$. The path of the connecting line between the sites of the dual lattice denotes the type of the  term in the expansion. The additional line denotes the part of the boundary (cf. Fig. \ref{boundary}). The sum is over all topologically equivalent contributions and over all possible locations ${\bf r}$ along the boundary line. 

What is clear from Eq.\eqref{pfb} is that there are distinct contributions in the perturbation series, coming from the corners of the boundary line. Having such a series expansion, we can derive a perturbation series for the entanglement entropy, using Eq. \eqref{enen}. We are going to restrict ourselves to terms up to order $O((v/u)^4)\sim O((\ln c^2)^4)$, so as to show the existence of a principle in this type of expansion, when a correlation length appears. The form of the series expansion depends on the number of degrees of freedom on the boundary $N_\Gamma$ and the number of corners $N_c$. Here, according to Fig. \ref{boundary}, $N_c=4$. We consider the physical limit $N_c\ll N_\Gamma$. At low orders in this expansion we find the result
\begin{eqnarray} 
S_A&=&\left[\ln2 -2\left(\frac{v}{u}\right)^4\right]N_\Gamma
\nonumber\\
&-&\ln2-\left[\frac{1}{2}\left(\frac{v}{u}\right)^2+\frac{5}{12}\left(\frac{v}{u}\right)^4\right]N_c+O\left(\left(\frac{v}{u}\right)^6\right)
\nonumber \\
&&
\label{8ven}
\end{eqnarray}  
From Eq.\eqref{8ven} it is clear that the correlation length effects amount to a renormalization of the term which is proportional to the length of the boundary and to a constant term which scales with the number of corners. On the other hand, the topological constant term remains invariant. It is apparent that the form of this result is a consequence of the structure of the expansion. Hence, the form of Eq.\eqref{8ven} will remain unchanged order by order in this expansion within its radius of convergence. Hence, the value of the topological entropy \emph{is a universal property of the topological phase}. In particular this form holds provided the size of the region is large compared to the correlation length. This requirement also applies to the distance between singular points (corners) on the boundary between the regions $A$ and $B$.

\section{The Quantum Dimer Model on the Triangular Lattice at the RK point }
\label{sec:tqdm}

In this Section, we show that the global topological degeneracy of the topological liquid phase of a quantum dimer model naturally leads to the existence of a topological term in the entanglement entropy. This term is directly related to the existence of  \emph{winding sectors} around the region of which the entanglement entropy is computed.

 We are going to focus on the properties of the ground state wave function of the triangular lattice quantum dimer model at the RK point\cite{rokhsar88}, even though our arguments hold more generally. 
These arguments hold for all topological ground state wave functions which have well defined loop representations and keep the assumptions made in Section~\ref{sec:entangle} intact. For example, in the general class of wave functions studied in Ref.~\onlinecite{freedman04}, the Temperley-Lieb algebra satisfied by the loops ensures that a loop cannot be broken into string segments, but can only be annihilated. This property, as we will show for the case of the dimer ground state wave function, enforces the existence of winding sectors around the region, whose entanglement entropy is calculated. 
 
  The dimer model on a triangular lattice at the RK point has the following ground-state wave function\cite{moessner2001}:
\begin{eqnarray}
\ket{G}=\frac{1}{\sqrt{Z}}\sum_{\cC}\ket{\cC}
\label{tdgs}
\end{eqnarray}
where $\cC$ labels configurations of hard-core dimers on the plane triangular lattice and the sum is over all configurations that may be connected through action of the ``flip term" (cf. Section \ref{sec:entangle} and Appendix \ref{sec:summary}).  This ground-state wave function is topologically ordered with 
a degeneracy that scales as $4^{\mathcal G}$, $\mathcal G$ being the genus of the surface on which the model is defined, and can be described by a $\mathbb{Z}_2$ gauge theory.  For example, on a torus, there are four equivalent topological sectors, each containing one of the four degenerate ground states.  In the following, we will consider $\eqref{tdgs}$ defined in one of these sectors.
\begin{figure}[htb]
\vspace{-1cm}
\subfigure[]{\includegraphics[width=1.65in]{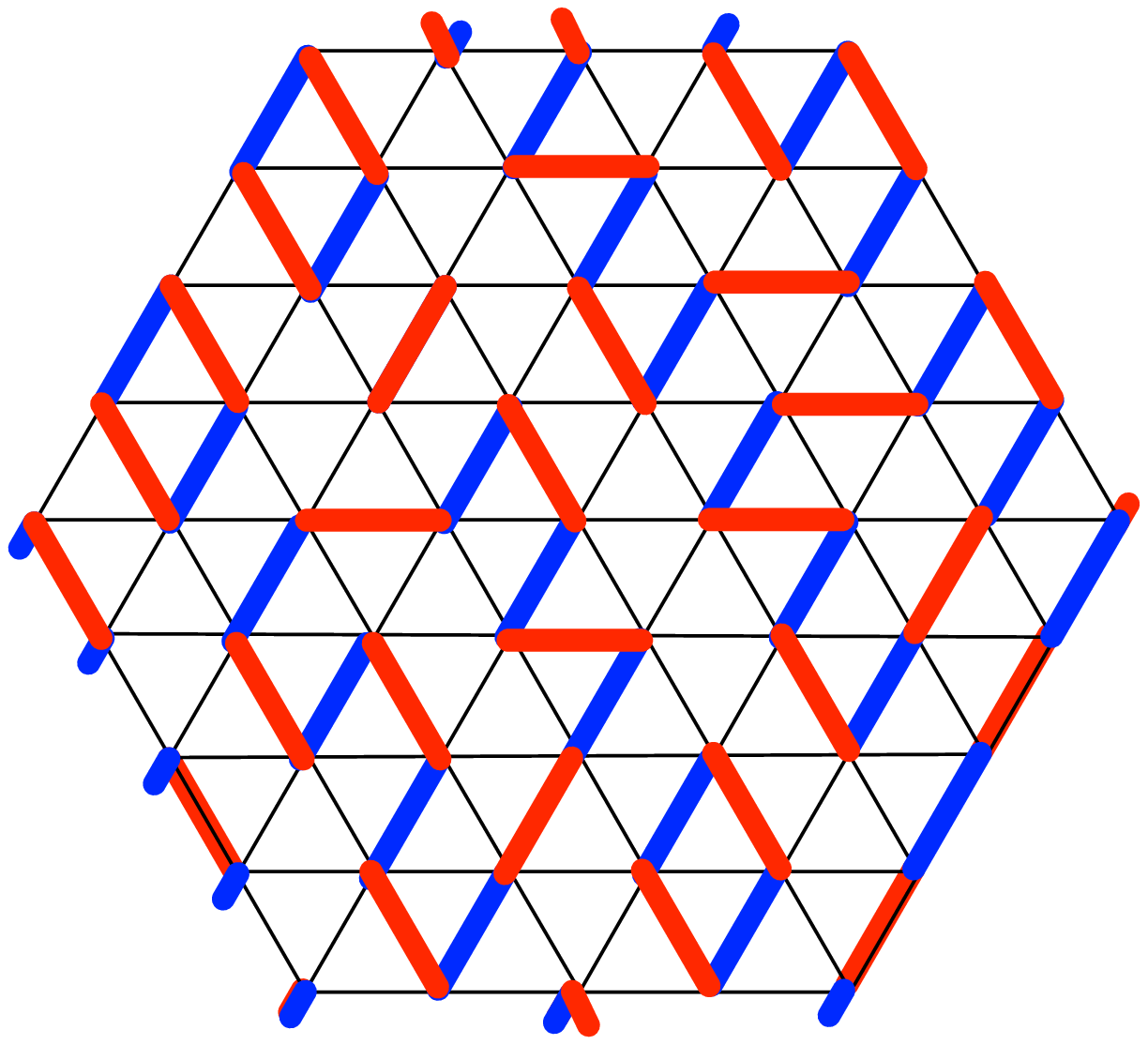}\label{transgraph} }
\subfigure[]{\includegraphics[width=1.65in]{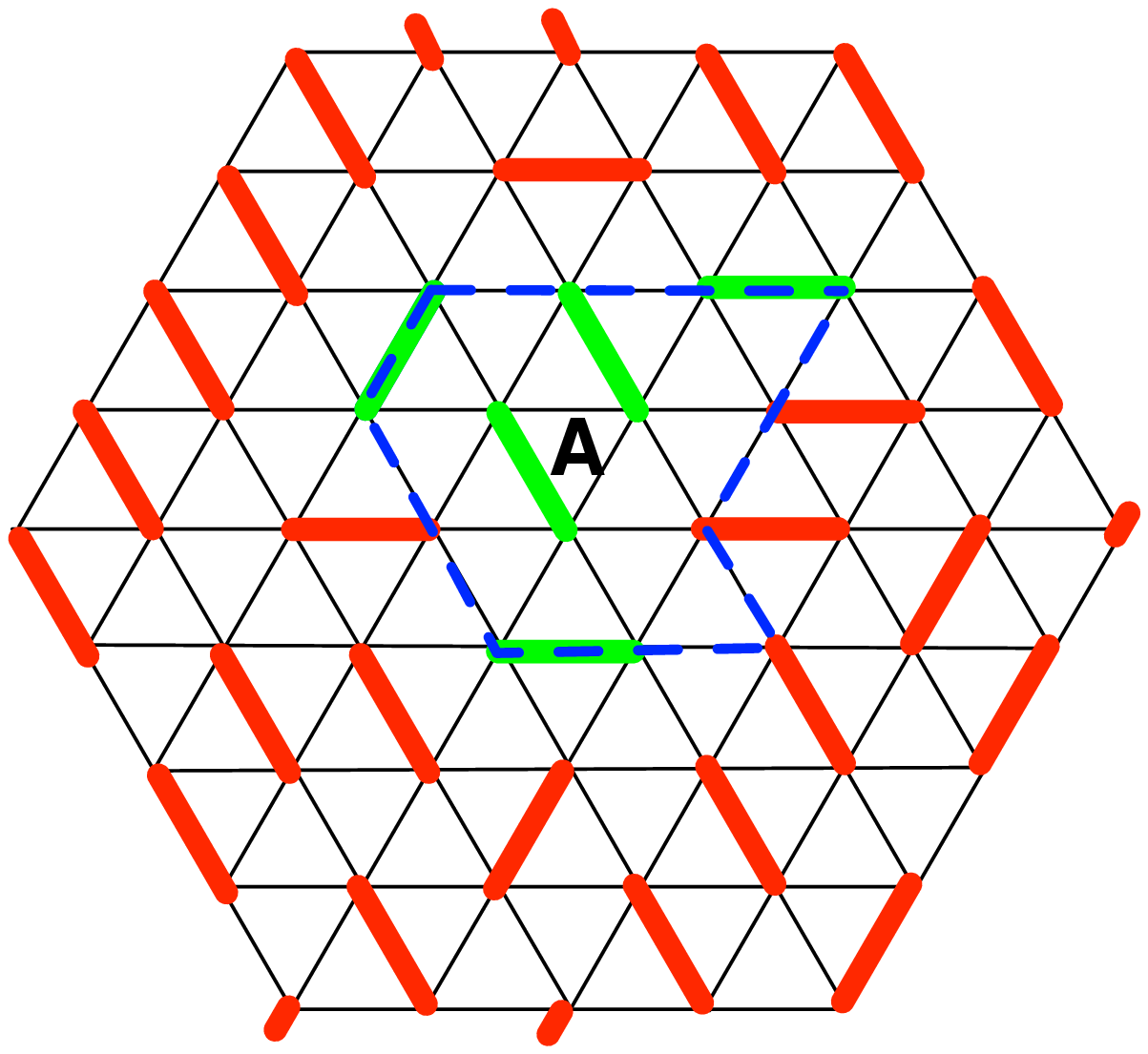}\label{tdbound}}
\vspace{-0.3cm}
\caption{(color online) (a) A part of the transition graph between a configuration $\cC$ (light(red)) and a reference columnar configuration $\cC_0$(dark(blue)). It is clear that the transition graph is composed typically of \emph{loops} and isolated \emph{links}. If the configuration $\cC$ is changed locally, then either the number of loops changes, adding or removing isolated links from the loops or the relative positions of the loops and the links change (without any shape change).(b) Definition of region $A$ with respect to region $B$. The light(green) dimers define the configuration in $A$ and the dark(red) ones the configuration in $B$. More generally, any link that lies on the boundary line or inside of it belongs to $A$. } 
\end{figure}
To elucidate more the structure of the ground state wave function $\eqref{tdgs}$, we 
consider the allowed transition graphs of the 
configurations $\ket{\cC}$ contained in $\eqref{tdgs}$, 
with a reference 
configuration $\ket{0}$, say one of the columnar states (cf. Fig.~\ref{transgraph}). The sum over all possible transition graphs is equal to the sum over all possible configurations. Now, given one transition graph, we consider all its possible transformations, which correspond to all possible gauge transformations in the corresponding gauge theory. A transition graph typically has two components, \emph{loops} and \emph{links}, but open lines are not allowed. The possible basic transformations classify to:
\begin{enumerate}
\item Flipping a loop to a set of links, 
\item Shifting a loop with respect to the links,
\item Combining two loops into a new one or breaking a loop into multiple loops,
\item Adding a link to a loop or the reverse.
\end{enumerate}
All of these irreducible transformations correspond to flips of different types of flippable dimer loops on the actual dimer configurations. Given this construction, it is easy to see that there are topological sectors in the configuration space of the system. If a loop exists that winds around the torus, for example, then no local transformation out of the four possible ones can lead to the removal of this loop, given that this loop is not flipped. If the loop is flipped, then the sector has changed. 

Now, we can apply these ideas to the concept of topological entropy. Imagine that we divide a closed space to regions $A$ and $B$ with a boundary loop $\Gamma$, where $A$ is located ``inside'' $B$, in the sense that a loop in $B$ which winds around $A$ is topologically non-trivial, thus making region $B$ not simply connected. In other words, by means of local operations the loop can only be 
deformed to become congruent with the boundary $\Gamma$  (cf. Fig.~\ref{tdbound}). Conversely, a loop in $B$ that does not wind around  region $A$ is contractible. For clarity, the boundary loop $\Gamma$ has been chosen in such a way that all dimers that lie on a link of the boundary belong to region A and are defined to be degrees of freedom of region $A$. 

Now, we consider the probability $p^{g}_\Gamma$, by fixing all the degrees of freedom on the boundary $\Gamma$. This can be done in this representation by specifying the dimers of region $B$ which intersect the boundary loop. Due to the hard-core constraint, all possible transformations included in the calculation of $p_{\Gamma}^g$ cannot change the fact that the same dimers intersect the boundary at the specified positions. We use the construction in terms of transition graphs but we choose the reference configuration in such a way that all the fixed dimers on the boundary match exactly. In this way, we prohibit the possibility that open lines are allowed in region $B$. Then, it is clear that the existence of region $A$ introduces a contribution of
\emph{winding sectors} of the configuration space (in loop language) in the calculation of the probabilities  $p^{g}_\Gamma$  of the boundary configurations. We select a single transition-graph loop winding around region $A$ and following the rules described, we can conclude that it cannot be destroyed by using the transformation rules 2 or 4, but it could by using rules 1 or 3. 
However, rule 3 can lead to the destruction of a non-contractible loop \emph{only if} it combines with another non-contractible loop around region $A$. Thus, if the number of non-contractible loops is even, then all these loops can be transformed to contractible ones by applying rule 3. On the other hand, if the number is odd, then the existence of a single non-contractible loop is unavoidable. Then, rule 1 or equivalently, a \emph{large transformation} around region $A$ is required to eliminate the loop. This fact leads us to the conclusion that  
\begin{eqnarray}
p^{g}_\Gamma=p^{+}_{\Gamma_g} + p^{-}_{\Gamma_g}
\label{fsec}
\end{eqnarray}
where $p_{\Gamma_g}^+$ and $p_{\Gamma_g}^-$ represent the probability $p^{g}_{\Gamma}$ with the constraint that there is an even or odd number, respectively, of non-contractible loops which \emph{wind} around region $A$. The emergence of this odd/even decomposition reveals the 
topological feature of the underlying gauge theory.

Now, we would like to ask whether there is any special relation of $p^{+}_{\Gamma_g}$ with $p^{-}_{\Gamma_g}$. If the correlation length of the system was zero, then the model would become one of the loop models considered in Ref.~\onlinecite{freedman04} with d-isotopy parameter $d=1$ (cf. Appendix~\ref{sec:summary}). In this case, it turns out that $p^{+}_{\Gamma_g}=p^{-}_{\Gamma_g}$ exactly\cite{freedman04}. In the same spirit, for topologically ordered ground-states of quantum dimer models, systems with finite correlation length, there is no reason why configurations in the odd sector should be favored or disfavored over those in the even sector, except possibly near the boundary $\Gamma$. Therefore, we conjecture that: 
\begin{eqnarray}
p_{\Gamma_g}^{+}=p^{-}_{\Gamma_g}\left(1+O(e^{-L_\Gamma/\xi})\right)
\label{seceq}
\end{eqnarray}
The reason why Eq.~\eqref{seceq} is expected to be true follows the arguments in Ref.~\onlinecite{wen90a}, for the degeneracy splitting of non-trivial topological sectors in a finite system. One can define, in general, large gauge transformation operators $\hat T$, around region $A$, which commute with the Hamiltonian and \emph{any other} physical observable operator. These operators map a state in one sector ($+$ or $-$) to a state in the other. In the dimer language, these operators flip a flippable dimer loop around region $A$. If we consider the operator $\hat S$ which has the property $\bra{G}\hat S\ket{G}=p_{\Gamma}^g$ then $\hat S$ is a sum of products of dimer density operators along the boundary contour $\Gamma$. Being in a topological phase, $\hat S$ commutes with $\hat T$, with corrections which depend on the tunneling probabilities between the $+$ and $-$ sectors. These probabilities, as shown in Ref.~\onlinecite{wen90a}, generally scale as $\sim e^{-L_\Gamma/\xi}$, leading ultimately to Eq.~\ref{seceq}.
 
The quantity $p_{\Gamma_g}^-$ can be estimated as follows: Given that the length of the boundary is $L_\Gamma$, the configuration that has the highest probability is the one which constitutes a flippable dimer loop along the boundary line. For a topological ground-state, the probability of such a flippable loop is just half the expectation value of the Wilson loop on the boundary $\Gamma$ and therefore is equal to\cite{simon1982} $e^{-c^{\Gamma}_0(\xi)L_{\Gamma}+c^{\Gamma}_1(\xi)}$, where $c_0(\xi)$ and $c_1(\xi)$ are correlation-length related effects which depend \emph{only} on the shape of the chosen boundary contour $\Gamma$.  Therefore:
\begin{eqnarray}
p_{\Gamma_g}^-= e^{-c_{0}^{\Gamma_g}(\xi)L_\Gamma+c_{1}^{\Gamma_g}(\xi)}
\end{eqnarray}
The argument holds for any $p_{\Gamma}^i$ so thus the entanglement entropy of region $A$ is:
\begin{eqnarray}
S_A &=& -\sum_{g} 2p_{\Gamma_g}^{-}\ln(2p_{\Gamma_g}^-)\nonumber\\
&=& c^0(\xi)L_\Gamma -c^1(\xi)-\ln2
\end{eqnarray}
where $c^{0,1}(\xi)=\sum_{g}p^{ }_{\Gamma_g} c^{0,1}_{\Gamma_g}(\xi)$.

This type of argument can be repeated for any similar topologically ordered quantum dimer model ground state wave function like the ones on the Kagome\cite{misguich02} and Fisher\cite{moessner03a} lattices. The existence of large  transformations  winding around region $A$ introduces non-trivial \emph{winding sectors} in the configuration space with a non-trivial contribution to the calculation of the entanglement entropy. 

\section{Defects, topological degeneracy, and topological entropy}
\label{sec:defects}

Defects, or equivalently violations of the local constraints of the model,  can be added in both dimer and vertex models in a similar way, given that there are known mappings between these classes. Such defects can be regarded as representing fluctuations of a matter field which, in the cases at hand, carries a  $\mathbb{Z}_2$ charge. We are going to focus on defects added in the ground-state wave function of a deconfined  topological phase at the Kitaev point in two different ways. Firstly, we examine the effect of such defects on the eight-vertex quantum model at the Kitaev point. In this model, in the representation where the link operator ``electric field'' $\hat\tau^{1}$ is diagonal, (electric) defects are vertices which have an imbalance of ingoing and outgoing arrows, namely vertices with 3 arrows in, 1 out or 3 out, 1 in. (See the notation and terminology of Section \ref{sec:q8v} and Appendix \ref{sec:summary}.) On the other hand, in the representation where the link operator (``gauge field'') $\hat\tau^{3}$ is diagonal, (magnetic) defects are defined by plaquettes which cannot resonate. Next, we examine the effects of virtual magnetic and electric charges on the Kitaev state by  looking at these effects in the problem of a $\mathbb{Z}_2$ gauge theory with matter. These two cases behave in clearly different ways which illuminate the problem of the stability of the topological phase.

It is clear that there are two physically distinct ways of adding such defects in the ground state wave function. Firstly, one can allow defects to be mobile, like a dilute gas. In this case, even though the ground-state wave function is connected perturbatively to the Kitaev wave function, local but non-perturbative terms have to be included in the Hamiltonian of the quantum model \cite{papanikolaou06}. Such a wave function corresponds to a state in which the $\mathbb{Z}_2$ matter field has ``condensed'', in the sense that free ($\mathbb{Z}_2$) gauge charges are proliferating in the ground state. It is a well known result\cite{fradkin-shenker79} that the state in which the matter field condenses is smoothly connected to a \emph{confining} state. In such a state, the loop configurations are essentially broken up into strings of finite length (the confinement scale). In such a state, the loops cannot explore the topology of the manifold on which the system is defined, or of the region of space being observed. Hence, such a state does not represent a topological phase. 

On the other hand, the quantum Hamiltonian can be perturbed by a term that favors the existence of such defects and charge neutrality should enforce the emergent defects (either magnetic or electric, see Appendix \ref{sec:summary}) to be created in pairs. In perturbation theory, the corrections to the ground state wave function contain even numbers of defects which are separated mutually by a distance typically of the order of the term in the expansion series. Only at very high orders in the perturbation series are defects more or less mobile, and an interpolation between the two limits of free and confined defects could be possible, if a phase transition is not present in between. Given that such a transition is typical in Ising gauge theories with matter fields in $2+1$ dimensions\cite{fradkin-shenker79}, one should expect that a similar transition happens in this case as well.

\subsection{Defect liquids}
\label{sec:defect-liquids}

We will now consider an extension of the Baxter wave function discussed in Section~\ref{sec:q8v} to include defects that violate the eight-vertex constraint. Namely, we will allow configurations with three arrows in and one out and vice versa. We will follow the notation and terminology of Section~\ref{sec:q8v} and Appendix \ref{sec:summary}.

More specifically, we will write the Kitaev wave function as a limit of a state which allows for the constraint to be violated at every vertex $\{ \textbf{r} \}$ of the lattice. In particular, if the Hilbert space is enlarged to include {\em all possible arrow configurations}, then the Kitaev wave function can be written as follows:
\begin{eqnarray}
\ket{G}=\sum_{\{\tau\}} e^{\sum_{\bf r}K_v(\tau_i \tau_j \tau_k \tau_l-1) }\ket{\{\tau\} }
\label{eq:wf-defect1}
\end{eqnarray}
where $i, j, k, l$ label the four links surrounding the vertex $\textbf{r}$, and we have denoted by $\tau_i$ the eigenvalue of the electric field operator $\tau^x_i$ ($\equiv \sigma^{1}_i$ of Section III) on link $i$ (and similarly with $j, k, l$). By direct examination of the wave function of Eq. \eqref{eq:wf-defect1} it is straightforward to see that 
in the limit $K_v\rightarrow\infty$, the exponential term enforces the eight-vertex constraint (even number of arrows in-even out) on each vertex of the lattice by setting to zero the amplitudes of all Ising (arrow) configurations which violate the constraint. It is instructive to generalize a little more this  wave function by adding an analogue of the classical \emph{polarization} field in the classical eight-vertex model:
\begin{eqnarray}
\ket{G}=\sum_{\{\tau\}} e^{\sum_{\bf r}K_v(\tau_i\tau_j\tau_k\tau_l-1) +K_\ell\sum_{i}\tau_i}\ket{\{\tau\} }
\label{isg2}
\end{eqnarray} 
When $K_{\ell}\gg1$, the only configurational state that has an appreciable amplitude is the one where $\tau_i=+1$ on every link $i$ of the lattice, a ``ferroelectric'' configuration. On the other hand, when $K_\ell\rightarrow0$, the usual Kitaev point is approached where the state is in a liquid phase. It is clear that a phase transition should separate these two limits and it does. 

It is immediate to see that the {\em norm} of the state $\ket{G}$ of Eq. \eqref{isg2} is simply the partition function of the $1+1$-dimensional Ising gauge theory coupled to a matter field. A well known duality transformation for Ising gauge theories in $1+1$ dimensions\cite{wegner71,balian75}, which maps the $1+1$ dimensional (in an Euclidean space-time lattice) Ising gauge theory with a matter field to a classical two-dimensional Ising model with an external magnetic field, can be applied here. Under this duality transformation, the norm of the ground state \eqref{isg2} is written in terms of Ising variables $s_i$, defined on the vertices of the lattice (instead of the links). Following closely Refs. \onlinecite{balian75} and \onlinecite{wegner71}, one easily finds that the norm of the ground state is:
\begin{eqnarray}
\braa{G}\ket{G}=\frac{1}{4}e^{-4N_sK_{v}}(\cosh(2K_v))^{N_s}(\cosh(K_\ell))^{2N_s}\nonumber\\
\sum_{\{s\}} e^{\sum_{<ij>}\beta^*s_is_j +h^*\sum_{i}s_i}
\label{isgdual}
\end{eqnarray}
where $i$ and $j$ now denote sites of the \emph{dual lattice}, and 
\begin{eqnarray}
\beta^*=-\frac{1}{2}\ln\tanh 2K_\ell,\hspace{0.5cm}h^*= -\frac{1}{2}\ln\tanh 2K_v
\label{dualrel}
\end{eqnarray}
It is important to notice that the factor $1/4$ in Eq.\eqref{isgdual} comes from the fact that on a torus, where the system is defined, periodic boundary conditions should be imposed on the direct lattice. This means that on the dual lattice, products of pairs of dual variables should equal on the boundary, leading to a double degeneracy for each of the directions of the torus.   

The amplitudes of the states that compose the ground state wave function follow the distribution of an Ising model in a magnetic field. In the defect-free limit $K_v\rightarrow\infty$, it follows from Eq.~\eqref{dualrel} that $h^*=0$. If $h^*\neq0$, then defects are present and independent from each other. So, the presence of mobile defects in the ground state of the quantum eight vertex model corresponds to a magnetic field which breaks the Ising symmetry of the weights of the wave function.
On the other hand, if we also set $K_\ell=0$ (which implies $\beta^*\rightarrow\infty$) to reach the Kitaev point, the weights of this wave function now have the form of the Gibbs weights of an Ising model at zero temperature and magnetic field. In this limit, there are just two \emph{dual} configurations which contribute a non-zero amplitude to the ground-state structure, the two fully ferromagnetic, up and down, Ising configurations. It is clear that this degeneracy in this dual formulation of the ground state wave function corresponds to the \emph{topological order} that characterizes this wave function.
\begin{figure}[tbh]
\subfigure[]{\includegraphics[width=0.18\textwidth]{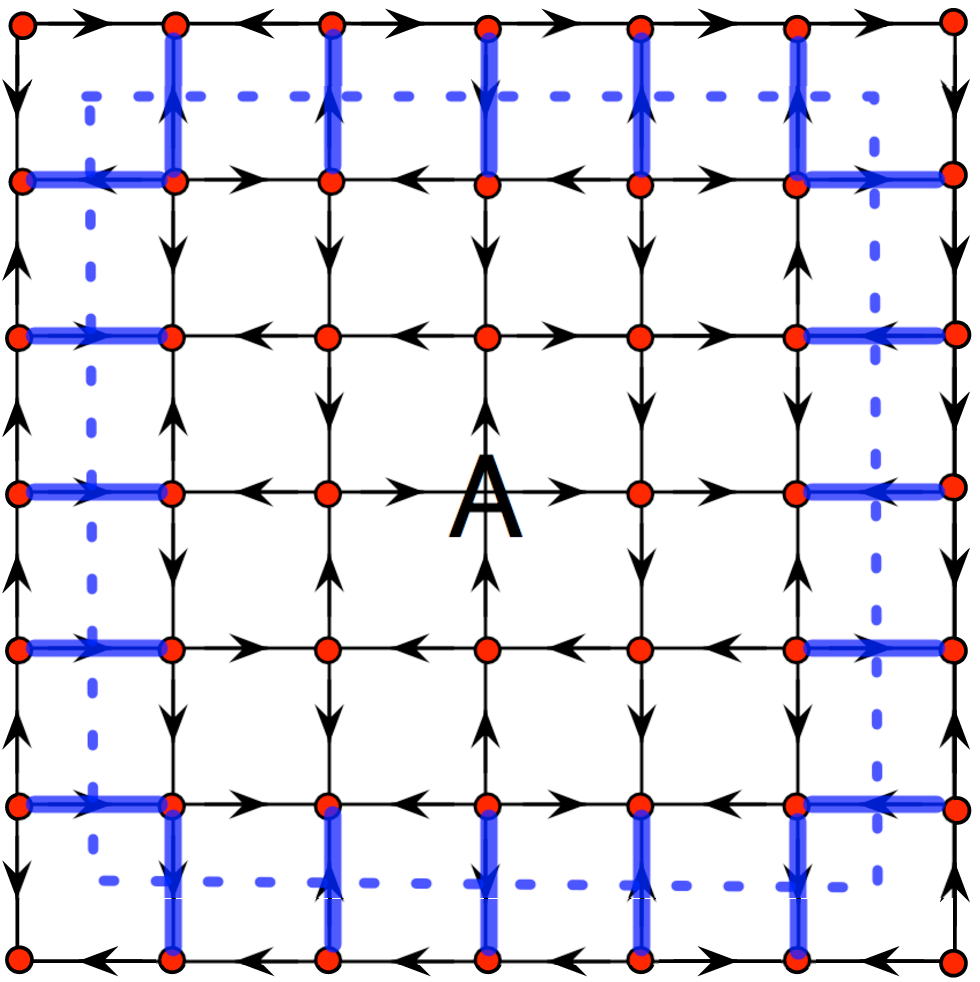}\label{dual}}
\subfigure[]{\includegraphics[width=0.18\textwidth]{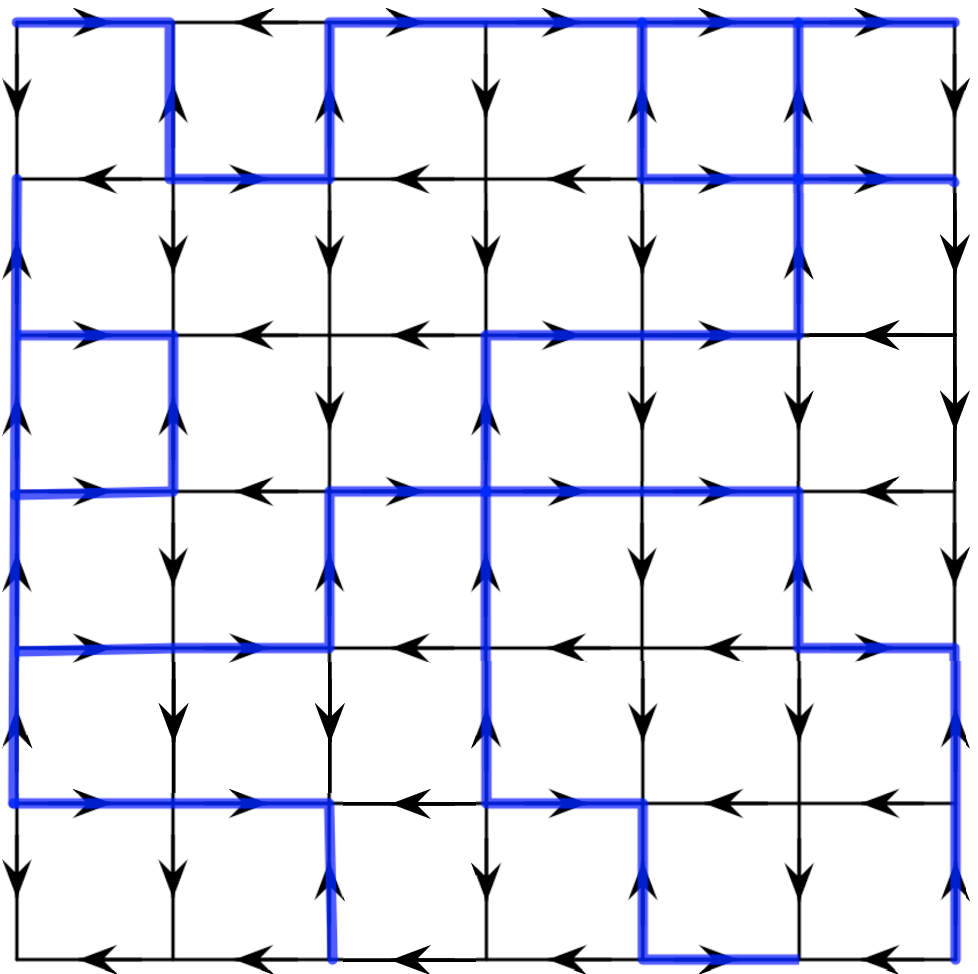}\label{string}}
\caption{(a) The positions of the arrows denote the direct lattice and the dots (located on the vertices) the corresponding dual lattice. The dashed line marks the boundary of region A and the thick bonds are the bonds on which the dual coupling $\beta^*$ has to reverse its sign when a correlation function $\left<\prod \tau \right>$, a dual Wilson loop, is to be computed. (b) The string representation of a typical eight-vertex configuration.}
\end{figure}

Let's consider the calculation of each $p_{\Gamma}^g$ as it is defined in Section~\ref{sec:q8v} and Fig.\ref{boundary}, {\t i.e.\/} the probability of a configuration on a loop $\Gamma$, the boundary of the region $A$ under observation for the calculation of the entanglement entropy. Using the methods of Ref. \onlinecite{kadanoff-ceva71}, an Ising variable $\tau_i$ in a correlation function maps under duality to the operator $e^{-2\beta^*s_{i^*}s_{j^*}}$, where $(i^*,j^*)$ represent the sites whose midpoint is the site $i$. The expectation value of such an operator in the dual formulation, in the limit $\beta^*\rightarrow\infty, h^*=0$, is clearly zero, because antiferromagnetic configurations are not present. In this way, 
\begin{eqnarray}
p_{\Gamma}^g &=& \frac{\bra{G}2^{-N_\Gamma}\prod_{i\in\Gamma}(1\pm e^{-2\beta^*s_{i^*}s_{j^*}})\ket{G}}{\braa{G}\ket{G}}
\label{eq:pgamma-g}
\end{eqnarray}
where the distribution of $\pm$'s in Eq.\eqref{eq:pgamma-g}, depends on the choice of the loop configuration $g_\Gamma$. In the relevant limit $\beta^*\rightarrow\infty$, all but two terms of the numerator's expansion are zero. The only non-zero terms are the first one $2^{-N_\Gamma}\braa{G}\ket{G}$ and the last one $2^{-N_{\Gamma}}\bra{G}\prod_{i\in\Gamma}e^{-2\beta^*s_{i^*}s_{j^*}}\ket{G}$. The latter term is non-zero and equal, in absolute value, to the former one because of the following reason: when all bonds on the (dual) loop $\Gamma$ are antiferromagnetic, they can all be satisfied without any energy penalty by having all spins in region $A$ being $+1$ and all spins outside region $A$ being $-1$, or the reverse. If the term is negative, then $p_{\Gamma}^g=0$, as it is expected because whenever $\prod\tau_x=-1$ around the loop, the eight-vertex constraint is violated. So, in the limit $\beta^*=\infty,h^*=0$ we have:
\begin{eqnarray}
p_{\Gamma}^g=2^{1-N_\Gamma}
\end{eqnarray}
for all the non-zero probabilities, and the known result in Eq.\eqref{kitaevpg} is recovered, as expected, leading to the entanglement entropy in Eq.\eqref{kitaevee}. 

If $\beta^*\gg1$ but finite($h^*=0$), then it is evident that the degeneracy between the two dual ferromagnetic states is present and exact until $\beta^{*}=\beta_c=\frac{1}{2}\ln(\sqrt2+1)$, where an Ising transition takes place to a paramagnetic state. For $\beta^* >\beta_c$, the topological entropy, using the same arguments as above, remains constant and equal to $\ln2$ until $\beta^* =\beta_c$ where it drops discontinuously to zero, since the exact two-fold degeneracy,  which led to the emergence of the topological entropy, is not present (see also Ref.~\onlinecite{castelnovo07b}). For $\beta^*>\beta_c$, the constant term of the entanglement entropy, computed from a low-temperature expansion of the dual system for $\beta^*\gg1$, contains exponentially small corrections($\sim e^{-C\beta^*}$), coming from antiferromagnetic fluctuations near the corners of the boundary between the ferromagnetic regions  $A$ and $B$(cf. Section~\ref{sec:q8v}). At the other limit, also similarly to our results in Section~\ref{sec:q8v}, for $\beta^*\rightarrow0$, only constant terms coming from the corners between the disordered paramagnetic regions $A,B$ are present, not related to the topology of the system.   

Given this simple way of calculating correlation functions, we can ask what is the effect of $K_v\gg1$ but finite, or equivalently $h^*\neq0$, introducing mobile defects in the wave function. Given that the exact degeneracy of the two dual ferromagnetic states is lifted when $h^*\neq0$, it is clear that the topological term should not be present in this case. This is actually the case in the limit $\beta^*\rightarrow\infty$
\begin{eqnarray}
p_{\Gamma}^g =2^{-N_\Gamma}\left(1 \pm \frac{\cosh[(N_A-N_B)h^*]}{\cosh(N_sh^*)}\right) 
\end{eqnarray}
where $N_A$ and $N_B$ are the numbers of dual sites in regions $A$ and $B$ respectively. The entanglement entropy is then,
\begin{eqnarray}
S_A=(\ln2)N_\Gamma +\left(\frac{\cosh[(N_A-N_B)h^*]}{\cosh(N_sh^*)}\right)^2+\cdots 
\end{eqnarray}
In the thermodynamic limit, clearly $S_A=(\ln2)N_\Gamma$, and the topological entropy has clearly been wiped out by the existence of mobile defects, {\it i.e.\/} $\gamma=0$. 

On the other hand, when $\beta^*\gg1$ but finite, clearly effects similar to the ones described in Section~\ref{sec:q8v} naturally appear. The reason is that on the direct lattice, the term proportional to $K_\ell$ in the amplitude weight of the wave function corresponds to a ``ferroelectric'' field in the classical eight vertex field, which, when weak, has similar effects to the ones described in Section~\ref{sec:q8v} (c.f. classic results from Baxter\cite{baxterbook}.) In terms of the dual formulation we presented here, it is clear that the exact Ising degeneracy, which leads to the topological entropy, persists \emph{for} $h^*=0$ \emph{until} $\beta^*=\beta^{*}_c$, where an Ising transition to a disordered state takes place.

\subsection{Perturbation theory and lifting of the topological degeneracy}
\label{sec:lifting}

The case $K_\ell\ll1,K_v\rightarrow\infty$ can be viewed also in another way. One can perturb the Kitaev Hamiltonian Eq.~\eqref{hk} with the following perturbation:
\begin{eqnarray}
\hat V=\lambda\sum_{\bf r}(\hat\tau^{x}_{\bf r}+1)
\label{perturb}
\end{eqnarray}
where $0<\lambda\ll t$. 
The effects of such perturbations have been studied recently, using numerical methods~\cite{hamma07, trebst07} who found, as expected, that the topological (deconfined) phase is stable. As it is discussed in detail in Appendix \ref{sec:summary}, this operator corresponds to an electric field in the language of the Ising gauge theory. If we consider the string representation of the eight-vertex configurations (cf. Fig.~\ref{string}), which can be defined by putting strings on links with an arrow pointing up or right, then clearly the term \eqref{perturb} does not favor the existence of strings on links of the lattice. In this representation, similarly to the discussion in Section~\ref{sec:tqdm}, topological sectors on a closed topological manifold, such as a torus, are distinguished by whether there is an even or odd number of strings winding each of the directions of the torus. Due to this fact, the perturbation Eq. \eqref{perturb} lifts the topological degeneracy, making the $(+,+)$ sector the most favorable, where the state with no strings belongs.

From another point of view, the operator $\tau_{\bf r}^x-1$ creates two \emph{magnetic defects} with an energy penalty $2t$  in the ground-state subspace, if one views the ground-state wave function in the representation where $\hat\tau^{z}_{\bf r}$ is diagonal. We will consider the states created by the application of products of these operators on the ground-state as the basis for applying perturbation theory. By applying formal Brillouin-Wigner perturbation theory, we take:
\begin{eqnarray}
\ket{G}&=&\ket{G_K}+\sum_{G^{(i)}_{\{\bf r\}}}  \frac{\bra{G^{(i)}_{\{\bf r\}}}V\Big| G \,\Big>}{E-\epsilon_{G^{(i)}_{\{\bf r\}}}}\ket{G^{(i)}_{\{\bf r\}}}\label{series}\\
&=& \ket{G_K}+c_1\ket{G_1}+\cdots
\end{eqnarray}
where $\ket{G_1}=\sum_{\{\bf r\}}\ket{G^{(1)}_{\{\bf r\}}} $  and $c_1=\frac{\lambda}{ 2t+\lambda}$. The state $\ket{G^{(1)}_{\{\bf r\}}}$ denotes the superposition of all possible configurations with $1$ magnetic defect at the location ${\bf r}$. Also, it is clear from the structure of the perturbation series that at a higher order n, the correction $\ket{G_n}$ emerges with a coefficient  $c_n$ of the order $O((\lambda/t)^n)$.

The entanglement entropy of the above wave function can be computed through our usual procedure, but only terms up to order $O(c_{1}^2)$ are going to be consistently correct. Higher orders in the entropy require the knowledge of higher orders in the perturbation series. Given that $\bra{G_K}\ket{G_1}=0$, we may consider the perturbed ground state as a pure state with unit amplitude (unnormalized) for 0-defect configurational states and $c_1$ for one-defect ones. Then, one can define a partition function (as in Section~\ref{sec:q8v}) and associated probabilities $p^{g}_\Gamma$, and proceed to the calculation of the entanglement entropy, as in Section~\ref{sec:q8v}. Following similar steps as before we have,
\begin{eqnarray}
S_A=(\ln2+c_{1}^2+O(\left(\lambda/t\right)^4))N_\Gamma -\ln2
\end{eqnarray}
It is clear that the topological term is robust under such low order perturbations and also, higher orders lead to corner effects, in a similar way as it is discussed in Section~\ref{sec:q8v}. This behavior has been verified numerically in Ref.~\onlinecite{hamma07}.  

From the form of the perturbation Eq. \eqref{perturb}, it is clear that the exact topological degeneracy of the \emph{winding sectors}, discussed in Section \ref{sec:tqdm} and related to the string representation ({cf. Fig.\ref{string}), is lifted. This happens because, if one considers the perturbation applied only on a loop around region $A$, then in the $+$ sector, the state with zero strings winding around region $A$ is favored, but in the $-$ sector, the state with $1$ string is favored with different energies. 

Given that the degeneracy is lifted, it is clear from our discussion throughout this paper that the entanglement entropy should acquire a non-trivial correction to the constant topological term, not related to the shape and type of the boundary, but directly related to the lifting of the winding sectors' degeneracy. In our example, this correction is identifiable. In order to show the principle,  let's keep just one relevant additional term in the wave function, which amounts to a set of terms which emerge in the perturbation series of Eq.~\eqref{series} to high orders: 
\begin{eqnarray}
\ket{\tilde G} = \ket{G_0}+c_{n_{\Gamma'}}\prod_{\Gamma'}(\tau_{r_{\Gamma'}}-1)\ket{G_0}
\end{eqnarray}
where $c_{n_{\Gamma'}}$ is a constant proportional to $(\lambda/t)^{n_{\Gamma'}}$ where $n_{\Gamma'}$ the length of a generic chosen loop which winds region $A$. The correction to the original wave function has the property that is non-zero whenever a string occupies all the links which lie on the loop $\Gamma'$. On the other hand, in the representation where $\hat\tau^{z}_{\bf r}$ is diagonal, the correction should be interpreted as the creation of a line of magnetic defects. Given that the winding sectors of the configurations surrounding region $A$ are characterized by even or odd number of strings, it is clear that this correction to the wave function has topological features. We can apply similar techniques as before to calculate the entanglement entropy, given there is no complexity with the definition of the boundary degrees of freedom (the loop $\Gamma'$ is chosen to be generically away from the boundary $\Gamma$ of region $A$). We can calculate the entanglement entropy, leading to the following results:
\begin{eqnarray}
p_{\Gamma}&=&\frac{2^{N_s-N_\Gamma } + (2c_{n_{\Gamma'}}+c_{n_{\Gamma'}}^2)2^{N_s-N_{\Gamma}-N_{\Gamma'}}}{2^{N_s-1}+(2c_{n_{\Gamma'}}+c_{n_{\Gamma'}}^2)2^{N_s-N_{\Gamma'}}}\\
&=&2^{1-N_\Gamma}(1-(2c_{n_{\Gamma'}}+c_{n_{\Gamma'}}^2)2^{-N_{\Gamma'}}+\cdots)
\end{eqnarray}
and the entanglement entropy is:
\begin{eqnarray}
&&S_A=(\ln2) N_\Gamma - \ln2 + 2^{-N_{\Gamma'}}(2c_{n_{\Gamma'}}+c_{n_{\Gamma'}}^2)+\cdots
\nonumber \\
&&
\end{eqnarray}
From the above expression it is clear that a correction to the topological term $\gamma=\ln 2$ appears, of the order $\sim(\lambda/(2t))^{N_{\Gamma'}}$. This is an exponentially small correction to the topological entropy which vanishes (exponentially fast) in the limit of a large region $A$, provided the perturbation theory used here is convergent. This is indeed correct for $\lambda < \lambda_c$, which is to say within the topological (deconfined) phase. On the other hand, these \emph{topological excitations}, when they proliferate ($\lambda\sim \lambda_c$), ultimately lead to the destruction of the topological entropy and, for a homogeneous system, to a global topological phase transition. As it is well known, the physics of this quantum critical point at $\lambda_c$ is equivalent (upon duality) to that of the quantum critical point of the $2+1$ dimensional Ising model in a transverse field, {\it a.k.a.\/} the classical three-dimensional Ising model. The quantity of interest here, the topological entropy, is thus related to the statistics of the proliferating domain walls of the 3D ising model at its critical point. The solution of this problem is still open.
  
Finally, we note that in this Subsection we considered only the effects of fluctuations on magnetic charges. The methods used here can be applied to the case of electric charges as well. Moreover, the well known \emph{self-duality} of the Ising gauge theory with Ising matter in $2+1$ dimensions\cite{balian75,fradkin-shenker79}, discussed in Appendix \ref{sec:summary}, implies that this result \emph{also applies for electric charges}, again within the domain of convergence of perturbation theory about the Kitaev (deconfined) point. This perturbation theory is well known to have a finite radius of convergence. We thus conclude that the topological entropy $\gamma=\ln 2$ is a property of the entire deconfined phase and not just of the Kitaev limit.

\section{Discussion and Concluding Remarks}
\label{sec:conclusions}

The calculations presented in this paper have implications for numerical attempts to determine the topological order in a ground-state wave function in the case of a finite correlation length.  The first issue to note is that there will be non-universal corrections to the subleading piece, as noted in Eq.~\eqref{sAgeneral}, and in the general case, the function $b(\xi)$ will not be known.  This makes the problem of extracting $\gamma$ more subtle than just calculating $S_A$ versus $L$ and then plotting the intercept.  

However, the observation in Section \ref{sec:q8v} that such non-universal corrections come from the corners of region $A$ suggests a way around this issue.  For example, on a square lattice and assuming the definitions of the boundary and its degrees of freedom of Section \ref{sec:q8v}, the entanglement entropy of a region with four corners will have the scaling form $S_A^{(4)}=a_4(\xi)L+b_4(\xi)+\dots$ where $b_4(\xi)=4c(\xi) -\gamma$ and $c(\xi)$ is the contribution of a single corner which will depend only on whether the corner lies on the direct or dual lattice.  The entropy $S_A^{(6)}$ of a region with six corners will have a similar form where $b_6(\xi)=6c(\xi)-\gamma$; here we have used the fact that a $90^\circ$ and $270^\circ$ corner give the same contribution.  Therefore, if we calculate $S_A^{(4)}$ and $S_A^{(6)}$ versus $L$ and determine the respective intercepts, then $\gamma$ may be extracted by:
\begin{equation}
\gamma = 2b_6(\xi) - 3b_4(\xi)
\label{subtract}
\end{equation}
This type of subtraction to eliminate the unknown effects of corners was one of the factors motivating the construction of Ref.~\onlinecite{kitaev2006}.  The present example is, perhaps, a simpler example of the same strategy. It is clear, though, that the application of such methods requires a precise understanding of the relevant degrees of freedom and the precise definition of the boundary shape and type between regions $A$ and $B$.

Additional simplifications arise for the special class of wave functions emphasized in this paper.  In Section \ref{sec:entangle}, it was shown that computing the entanglement entropies of wave functions whose normalizations resemble partition functions of classical statistical mechanical systems, amounts to computing probabilities of boundary configurations of the same classical systems.  Such probabilistic quantities are convenient to compute using classical approximate techniques such as Monte Carlo simulation.  The advantage of such techniques is that it becomes possible to investigate system sizes large enough that intrinsic finite size effects, such as corners interacting with each other or
perturbative mixing of topological sectors as discussed in Section \ref{sec:lifting}, are no longer an issue.

Moreover, by working with the reduced quantities mentioned in Eqs.~\eqref{suben} and \eqref{conjform},
the problem further reduces to one of computing expectation values of non-local loop operators of varying loop shapes.  The topological entropy may be extracted using the strategy of Eq.~\eqref{subtract}, for example, though care should be taken to choose the loop operators such that the different corners of the loop have similar immediate environments so that the subtractions go through.  

This viewpoint provides a route to answering certain unresolved questions about Fig.~\ref{8vpd} away from the Kitaev point such as the mechanism by which the topological order is lost along the critical lines to give way to the ordered ``antiferroelectric'' phases.  Such a study, which further connects the present work to ideas discussed in Ref.~\onlinecite{fradkin06}, is a natural topic for further investigation. 

In this paper, we tried to elucidate the topological features of the entanglement entropy of topologically ordered ground-state wave functions with finite correlation length. Firstly, by expanding generally the entanglement entropy in terms of appropriately defined boundary probabilities, we suggested that reduced quantities, directly related to the usual von Neumann entanglement entropy, contain the same topological information as the entanglement entropy. Then, we showed explicitly in the context of the quantum eight-vertex model, that the effects from the existence of a finite correlation length amount to contributions to the linear part and non-topological constant pieces coming from non-smooth parts of the boundary. Then, with reference to the topologically ordered ground-state wave function of the quantum dimer model at the RK point, we showed that the concept of topological entropy is connected to the existence of ficticious sectors in the calculation of the entanglement entropy, allowing for the identification of the topological order of the state. Finally, we considered the effects of topological defects on the entanglement entropy of the quantum eight vertex model near the Kitaev point. We firstly considered the non-perturbative regime of a defect liquid, where defects are mobile and cause the topological entropy to vanish. Secondly, we showed that if defects are slightly favored in the quantum Hamiltonian, the topological entropy is a topological invariant and effects related to the shape and type of the boundary are only allowed. Finally, along these lines, we identified the terms in the perturbation expansion which render topological corrections to the topological entropy and should ultimately lead to a global topological phase transition, if they proliferated.

Note: As this work was being completed, we became aware of the work by Castelnovo and Chamon\cite{castelnovo07b}, who independently, and among other questions, considered the wavefunction in Eq.~\eqref{isg2} in the limit $K_{v}\rightarrow\infty$ and studied its topological entropy as a function of $K_\ell$ in the context of an investigation of the stability of a topological phase. We thank these authors for communicating their results with us before publication.

\begin{acknowledgments}
We thank Claudio Castelnovo, Claudio Chamon, Paul Fendley, Duncan Haldane, Michael Levin, Joel Moore, Chetan Nayak, John Preskill, Rahul Roy and Kirill Shtengel  for many discussions. This work was supported in part by the National Science Foundation through the grant NSF DMR 0442537, and by the U.S. Department of Energy, Division of Materials Sciences under Award DEFG02-91ER45439, through the Frederick Seitz Materials Research Laboratory at 
the University of Illinois at Urbana-Champaign.  
\end{acknowledgments}
  
\appendix
\section{A summary of relevant $\mathbb{Z}_2$ gauge theory results}
\label{sec:summary}

In this Appendix we will review a number of standard and well known results from the gauge theory literature in the context of its applications to topological phases. Much of what we discuss here was developed and reviewed extensively in Refs.~\onlinecite{kogut75}, \onlinecite{fradkin78}, \onlinecite{kogut79} and \onlinecite{fradkin-shenker79}.  The topological nature of deconfined phases of discrete gauge theories was emphasized in Refs. \onlinecite{krauss89} and \onlinecite{preskill90} and its role in topological mechanisms for quantum computing was first formulated by Kitaev\cite{kitaev97} (for a recent review see \onlinecite{dassarma07}.) Here we will focus on the deconfined phase of $\mathbb{Z}_2$ gauge theory with matter\cite{wegner71,balian75,fradkin78,fradkin-shenker79} which is the simplest example of a topological phase. Some aspects of this problem were revisited recently in Ref.~\onlinecite{hastings2005}.

Let us consider a $\mathbb{Z}_2$ gauge theory with matter fields on a  $2+1$-dimensional square lattice in the Hamiltonian formulation. As usual the $\mathbb{Z}_2$ Ising gauge fields live on the links of the square lattice. We denote by the Pauli matrix $\tau^j_z(\bf r)$, the gauge field on the link $(\bf r, {\bf r}+ {\bf e_j})$ of the square lattice (with $j=1,2$ denoting the two spacial directions). The (Ising) matter field resides on the sites $\bf r$ of the square lattice and is denoted by the Pauli matrix $\sigma_z(\bf r)$. The quantum Hamiltonian is
\begin{eqnarray}
H&=& -g \sum_{{\bf r},j=1,2} \tau_x^j ({\bf r}) -\sum_{\bf r} \sigma_x({\bf r})\nonumber \\
&&-\sum_{\bf r} \tau_z^1({\bf r}) \tau_z^2({\bf r}+{\bf e}_1)\tau_z^1({\bf r}+{\bf e}_2)\tau_z^2({\bf r})\nonumber \\
&&-\lambda \sum_{{\bf r}, j=1,2} \sigma_z({\bf r}) \tau^j_z({\bf r}) \sigma_z({\bf r}+{\bf e}_j)
\label{eq:ising-matter}
\end{eqnarray}
where $g$ and $\lambda$ are two coupling constants. The first term in Eq.\eqref{eq:ising-matter} is a gauge field kinetic energy (electric field-like) term, the second is a matter kinetic energy (a transverse field), the third term acts on plaquettes and is a gauge potential energy (magnetic flux) term, and finally the last term minimally couples the matter and gauge fields.

This quantum Hamiltonian has a local $\mathbb{Z}_2$ gauge symmetry. The generators  $G({\bf r})$ of local $\mathbb{Z}_2$ gauge transformations are,
\begin{equation}
G({\bf r})\equiv \sigma_x({\bf r}) \tau^1_x({\bf r})\tau^1_x({\bf r}-{\bf e}_1)
\tau^2_x({\bf r})\tau^2_x({\bf r}-{\bf e}_2)
\label{eq:gauge-generators}
\end{equation}
These are Ising operators
\begin{equation}
G({\bf r})^2=1
\end{equation}
which commute with each other 
\begin{equation}
\left[G({\bf r}), G({\bf r}^\prime)\right]=0
\end{equation}
and with the Hamiltonian
\begin{equation}
\left[G({\bf r}), H\right]=0
\end{equation}
The physical Hilbert space are the gauge invariant states $\ket{\textrm{Phys}}$, 
\begin{equation}
G({\bf r}) \ket{\textrm{Phys}}=\ket{\textrm{Phys}}, \quad \forall \; \textbf{r}
\label{eq:gauge-invariance}
\end{equation}
{\it i.e.\/} the physical states invariant under arbitrary local time-independent gauge transformations.
In the basis in which the gauge field $\tau^j_z({\bf r})$ and the matter field $\sigma_z({\bf r})$ are diagonal, the action of the local gauge transformations $G({\bf r})$ is to flip the sign of the matter field $\sigma_z$ at site ${\bf r}$ and of the gauge fields $\tau^j_z$ on the four surrounding links. 
The simpler physical states of this theory can be qualitatively described as {\em magnetic charges}, {\it i.e.\/} plaquettes where the magnetic flux term is $-1$, and {\em electric charges}, {\it i.e.\/} sites $\bf r$  where $\sigma_x({\bf r})=-1$.

A local gauge symmetry cannot be spontaneously broken.\cite{elitzur75} However it is possible to use the gauge invariance of the theory to fix the gauge. In this case it is possible to use the action of the gauge generators $G({\bf r})$ to fix the gauge,
\begin{equation}
\sigma_z({\bf r})=1
\label{eq:unitary}
\end{equation}
globally and completely, on {\em all sites} $\bf r$ of the square lattice for {\em any} boundary conditions (a disk, a torus, etc.) This is the unitary-London  gauge. In a general gauge theory, the gauge in which the phase of the matter field is set to zero is known as the unitary gauge, and in the theory of superconductivity it is known as the London gauge. In this Ising theory, this gauge fixes the matter degrees of freedom completely.

{\em In this gauge} the Hamiltonian takes the form
\begin{eqnarray}
H&=& -g \sum_{{\bf r},j=1,2} \tau_x^j ({\bf r}) \nonumber \\
&&-\sum_{\bf r} \tau_x^1({\bf r})
\tau_x^1({\bf r}-{\bf e}_1)\tau_x^2({\bf r})\tau_x^2({\bf r}-{\bf e}_2)\nonumber \\
&&-\sum_{\bf r} \tau_z^1({\bf r}) \tau_z^2({\bf r}+{\bf e}_1)\tau_z^1({\bf r}+{\bf e}_2)\tau_z^2({\bf r})\nonumber \\
&&-\lambda \sum_{{\bf r}, j=1,2}  \tau^j_z({\bf r}) 
\label{eq:H-unitary}
\end{eqnarray}
In this gauge the local symmetry is absent (which is natural since we have fixed the gauge completely). 

Ising gauge theories, with and without matter, have well known {\em duality} mappings,\cite{wegner71,balian75,fradkin78} which are straightforward generalizations of the Kramers-Wannier duality of the 2D classical Ising model. In short, in $1+1$ dimensions Ising models are self-dual, while in $3+1$ dimensions Ising gauge theories are self dual instead. In $2+1$ dimensions Ising models are dual to gauge theories, and, in particular, quantum Ising gauge theories coupled to Ising matter fields are {\em self dual} in $2+1$ dimensions. In the context of the unitary-gauge Hamiltonian, Eq.\eqref{eq:H-unitary}, self duality of the theory amounts to the simple mapping $g \leftrightarrow \lambda$, which requires the phase diagram to be invariant under this symmetry, with the caveat that the electric and magnetic properties of the states are exchanged under this duality symmetry.

The phase diagram of this theory is well known.\cite{fradkin-shenker79} If the dimensionality $d$ of {\em space} is greater than $1$, it has two distinct phases: a) a confined phase, smoothly connected from a so-called Higgs phase (from which it cannot be distinguished), and b) a free charge (or Coulomb) phase. 
The confinement-Higgs phase has been extensively discussed in the literature\cite{fradkin-shenker79} and we will not discuss it here. It is a massive phase with a unique ground state and a gauge-invariant spectrum of massive magnetic charges and no electric charges. It occupies the strong coupling sector of the phase diagram, where the gauge coupling $g$ and/or the matter coupling $\lambda$ are large.

The deconfined or free charge (Coulomb) phase occurs when both the gauge and matter couplings are weak, $0\leq g <g_c$ and $0\leq \lambda <\lambda_c$, where $g_c$ and $\lambda_c$  are critical couplings. In this phase the spectrum contains finite energy states which carry the $\mathbb{Z}_2$ charge\cite{fradkin-shenker79}. The deconfined phase is controlled by the infrared stable fixed point at $g=\lambda=0$ which describes a gauge theory in the extreme deconfined limit and matter fields which are infinitely heavy. At this fixed point the Hamiltonian takes the simpler form  
\begin{eqnarray}
H&=&  -\sum_{\bf r} \tau_x^1({\bf r})
\tau_x^1({\bf r}-{\bf e}_1)\tau_x^2({\bf r})\tau_x^2({\bf r}-{\bf e}_2)\nonumber \\
&&-\sum_{\bf r} \tau_z^1({\bf r}) \tau_z^2({\bf r}+{\bf e}_1)\tau_z^1({\bf r}+{\bf e}_2)\tau_z^2({\bf r})
\label{eq:H-toric}
\end{eqnarray}
known as  Kitaev's toric code\cite{kitaev97} Hamiltonian.
It consists of the sum of two mutually commuting sets of operators: the first term is the generator of time-independent gauge transformations of the pure gauge theory and it is the product of four  $\tau_x$ operators residing on the links emanating from the site (or vertex) $\bf r$. The second term, in which the operators $\tau_z$ reside on the links of the plaquette labeled by the site $\bf r$, is the actual Hamiltonian. Clearly, in this limit,  the eigenstates of the first term are states with or without $\mathbb{Z}_2$ electric charges (present if this operator takes the value $-1$) while the eigenstates of the second (plaquette) term are states with or without $\mathbb{Z}_2$ magnetic charges (present if this term takes the value $-1$). In this extreme deconfined limit, the ground state has neither electric nor magnetic charges.

We end this Appendix with a description of two alternative, dual, descriptions of the states in the deconfined phase. 
\begin{enumerate}
\item
The first approach is the standard description of the states in a gauge theory in terms of the eigenstates of the vector potential which in this $\mathbb{Z}_2$ theory are the eigenstates of the operators $\tau^j_z(\bf r)$, which are not gauge-invariant states. At this stage there are two options on how to proceed. One option, which is indeed the standard approach, consists of fixing the gauge. This works provided it is possible to fix the gauge completely, something which is not possible in many gauges on manifolds with non trivial topology such as the torus. The alternative is to define gauge invariant states as the linear superposition obtained by the action of the gauge group $\mathbb{Z}_2$ on some reference gauge-fixed state, {\it i.e.\/} by all possible actions with the gauge generators $G(\bf r)$. In this approach it is essential to account for the action of  the {\em large gauge transformations}, transformations in which the action of the gauge generators wraps around the non contractible closed contours $\Gamma_k$ of the space manifold, {\it e.g.\/} $k=1,2$ for the torus. This effectively splits the configurations of gauge fields into classes each labeled by the expectation value of the (gauge-invariant) Wilson loop operators along each non contractible contour $\Gamma_j$, {\it i.e.\/} $W_j=\prod_{\textbf{r} \in \Gamma_j} \tau^j_z(\bf r)$. On the other hand, the generators of large gauge transformations are the operators that create Dirac strings along the same non contractible contours $\widetilde \Gamma_j$, the closed contours on  the {\em dual} lattice. These Dirac string, dual, generators are denoted by ${\widetilde W}_j= \prod_{\textbf{r} \in \widetilde \Gamma_j} \tau^j_x(\bf r)$. These mutually dual sets of Wilson loops obey the simple
algebra,\cite{freedman04}
\begin{equation}
\left[W_j, \widetilde W_j\right]=0, \quad \left\{W_j, \widetilde W_k\right\}=0, \; \textrm{for} \; j \neq k
\label{thooft}
\end{equation}
\item
The second alternative consists in working directly instead with gauge-invariant states, as discussed extensively by Freedman and coworkers.\cite{freedman04} This is the approach that we followed in this paper. In this approach, the space of states are eigenstates of the Wilson loop operators on {\em arbitrary} closed contours $\Gamma$, contractible or not. In addition, one uses the eigenstates of the electric field link operators  $\tau_x^j$, which are gauge invariant by construction. A simple picture of the space of states in this basis is obtained by mapping it to the set of closed loop configurations on the square lattice: a link belongs to a loop if $\tau_x=-1$ on that link and it does not otherwise. The magnetic plaquette term of the Hamiltonian acts on neighboring loops and reconnects their strands (thus changing the loop configurations). When acting on an elementary closed loop configuration, which occupies just one plaquette,  it annihilates it with weight $d=1$. This algebra furnishes a representation of  the generator of a Temperley-Lieb algebra with d-isotopy parameter $d=1$, as discussed in Ref. \onlinecite{freedman04}. In this picture, the ground state in the extreme deconfined (Kitaev) limit is an eigenstate of the magnetic plaquette term and it is thus the state with equal amplitude superposition of all loop configurations, discussed elsewhere in this paper. It is a peculiarity of the $\mathbb{Z}_2$ gauge theory that the loop and gauge field representations are equivalent, and in fact isomorphic to each other.
\end{enumerate}

\end{document}